\documentclass[aps, pra, a4paper, showpacs,10pt, twocolumn, superscriptaddress]{revtex4-1}
\usepackage{bbm, amsmath, amssymb, amsthm, bm,textcomp, nicefrac,geometry}
\usepackage{graphicx,epstopdf,color,verbatim,enumitem, mathtools}
\usepackage[dvipsnames]{xcolor}
\usepackage{ dsfont }

\definecolor{myurlcolor}{rgb}{0,0,0.7}
\definecolor{myrefcolor}{rgb}{0.8,0,0}
\usepackage[unicode=true,pdfusetitle, bookmarks=false,bookmarksnumbered=false,
bookmarksopen=false, breaklinks=false,pdfborder={0 0 0},backref=false,
colorlinks=true, linkcolor=myrefcolor,citecolor=myurlcolor,urlcolor=myurlcolor]
{hyperref}

\graphicspath{ {./figs/} }

\theoremstyle{definition}

\newcommand{\eins}{\mathbbm{1}}

\newcommand{\ket}[1]{\left|#1\right\rangle}
\newcommand{\bra}[1]{\left\langle #1\right|}
\newcommand{\bracket}[2]{\left\langle #1|#2\right\rangle}
\newcommand{\coh}[2]{\left|#1\rangle \!\langle#2\right|}
\newcommand{\prjct}[1]{\left|#1\right\rangle\!\left\langle #1\right|}

\renewcommand{\t}[1]{\textrm{#1}}

\newcommand{\be}{\begin{equation}}
\newcommand{\ee}{\end{equation}}
\newcommand{\bea}{\begin{align}}
\newcommand{\eea}{\end{align}}
\newcommand{\cE}{\mathcal{E}}
\newcommand{\cL}{\mathcal{L}}
\newcommand{\sS}{\mathbb{S}}
\newcommand{\tr}{\textrm{tr}}
\newcommand\ii{\mathrm{i}}
\newcommand{\mc}[1]{\mathcal{#1}}
\newcommand{\expect}[1]{\langle #1 \rangle}
\newcommand{\dd}{\mathrm{d}}

\DeclarePairedDelimiter\floor{\lfloor}{\rfloor}

\begin{document}
\title{General measure for macroscopic quantum states beyond ``dead and alive''}

\author{Pavel Sekatski}
\affiliation{Institut
f\"ur Theoretische Physik, Universit\"at Innsbruck, Technikerstra\ss e 21a, 6020 Innsbruck, Austria}
\author{Benjamin Yadin}
\affiliation{Department of Atomic and Laser Physics, University of Oxford, Parks Road, Oxford OX1 3PU, United Kingdom}
\author{Marc-Olivier Renou}
\affiliation{Group of Applied Physics, University of Geneva, 1211 Geneva, Switzerland}
\author{Wolfgang D\"ur}
\affiliation{Institut
f\"ur Theoretische Physik, Universit\"at Innsbruck, Technikerstra\ss e 21a, 6020 Innsbruck, Austria}
\author{Nicolas Gisin}
\affiliation{Group of Applied Physics, University of Geneva, 1211 Geneva, Switzerland}
\author{Florian Fr\"owis}
\affiliation{Group of Applied Physics, University of Geneva, 1211 Geneva, Switzerland}
\begin{abstract}
  We consider the characterization of quantum superposition states beyond the pattern ``dead and alive''. We propose a measure that is applicable to superpositions of multiple macroscopically distinct states, superpositions with different weights as well as mixed states. The measure is based on the mutual information to characterize the distinguishability between multiple superposition states. This allows us to overcome limitations of previous proposals, and to bridge the gap between general measures for macroscopic quantumness and measures for Schr\"odinger-cat type superpositions. We discuss a number of relevant examples, provide an alternative definition using basis-dependent quantum discord and reveal connections to other proposals in the literature. Finally, we also show the connection between the size of quantum states as quantified by our measure  and their vulnerability to noise. 
\end{abstract}
\date{\today}

\maketitle

\section{Introduction}
\label{sec:introduction}

When going from single microscopic particles to composite systems with many degrees of freedom, quantum mechanics shows enormous complexity. Genuine quantum features such as entanglement or nonlocality fall into several subclasses and notions such as ``maximally entangled state'' can not be generalized in a straightforward way. A full characterization of meso- or even macroscopic quantum systems seems to be out of reach, not only for practical reasons.
However, one can identify global properties in such systems that are only mildly influenced by microscopic details. One such aspect is the macroscopic quantumness in large quantum systems. 

An historic example that has played an important role is the so-called Schr\"odinger-cat state \cite{Schrodinger_Present_1935}. The pictorial idea of Schr\"odinger's cat in a macroscopic superposition of dead and alive and entangled with a radioactive atom is easy to grasp. However, as first emphasized by Leggett \cite{Leggett_Macroscopic_1980}, the so-called  macroscopic distinctness of the two superposed components $| \mathrm{alive} \rangle + \left| \mathrm{dead} \right\rangle $ is a particularity not present in any quantum effect brought to macroscopic scales. For a counterexample, Leggett mentions superconductivity on visible scales. In order to further elaborate on this difference, several proposals to formalize the concept of macroscopic distinctness based on the ``dead and alive'' structure of a quantum state have been put forward \cite{Leggett_Testing_2002,Dur_Effective_2002,Korsbakken_Measurement-based_2007,Marquardt_Measuring_2008,Sekatski_Size_2014,Laghaout_Assessments_2015}. For instance, the redundancy of information encoding in subparts of the system (like in the cells of the biological cat) \cite{Korsbakken_Measurement-based_2007}, or the distance measured in units of ``microscopic steps'' \cite{Marquardt_Measuring_2008} have been suggested. Even though these approaches are conceptually appealing, they suffer from some shortcomings. A general pure state does not have a Schr\"odinger-cat like structure, and, though one can always try to find a decomposition of a state into ``dead and alive'', such a decomposition is never unique. Even in the case where a natural choice seems to exist, this may not automatically lead to the maximal result \footnote{See, for instance, the examples discussed in \cite{Korsbakken_Measurement-based_2007,Marquardt_Measuring_2008}}. In addition, extensions to superposition with different weights, or to mixed states are not straightforward. This limits the proposals to analyze ideal situations, while experimental data is difficult to interpret.

Other measures are directly formulated for arbitrary quantum states \cite{Leggett_Macroscopic_1980,Bjork_Size_2004,Shimizu_Detection_2005,Lee_Quantification_2011,Frowis_Measures_2012,Yadin_Quantum_2015}. Some of them are based on a pre-chosen observable of the system and define generalized notions of ``macroscopically distinct'' as the spread of the wave function in the spectrum. The variance of this observable for pure states is closely connected to some proposals \cite{Bjork_Size_2004,Shimizu_Detection_2005,Lee_Quantification_2011,Frowis_Measures_2012}. The more general approaches are however sometimes criticized to lack the conceptual beauty and clear physical intuition (as given by the distinctness of the two components for the other measures).

In this paper, we close the gap between these two basic approaches. We propose a measure that is applicable to superpositions of multiple states with unequal weights and is readily extendable to mixed states, thereby overcoming the shortcomings of previous proposals. We start with the intuition that $| \mathrm{alive} \rangle $ and $| \mathrm{dead} \rangle $ are macroscopically distinct if the two states can be distinguished by ``classical'' detectors \cite{Sekatski_Size_2014}, i.e. detectors that do not in general completely collapse the system into perfectly orthogonal states upon measurement but only weakly disturb the system. Needless to say that such detectors also do not perfectly extract information about any state of the system, hence they are said to have a limited resolution precision or resolution.
Considering general pure states $| \Psi \rangle $ without specifying a subdivision into two branches, we quantify how much information about $| \Psi \rangle $ can be extracted by such detectors. The informative content is measured with the mutual information between system and measurement apparatus in the relevant bases. Then, we attribute an effective size to $| \Psi \rangle $ as the robustness of this information with respect to the detector's resolution. The concept of macroscopic distinctness is hence formalized as ``macroscopically extractable information''. 
This idea is generalized to mixed states via a convex roof construction (i.e., considering the ``worst case''). In contrast to \cite{Laghaout_Assessments_2015}, which follows a similar approach using pairwise distinguishability, using the mutual information ensures a holistic treatment, and allows for a connection between different approaches.  

The paper is summarized as follows. In Sec.~\ref{sec:macr-inform-capac}, we formalize the intuitive idea and define our measure for pure states. We discuss a paradigmatic example of multiple superposition states and establish a connection to the variance. In Sec.~\ref{sec:mixed-states-convex}, we present an extension of our measure to mixed states using a convex-roof construction, and illustrate it with a simple example. In Sec.~\ref{sec:altern-comp} we introduce alternative ways to formalize our intuition. For one of them, we use the basis-dependent discord. We discuss implications and connections between the alternatives and to other proposals from the literature. In Sec.~\ref{sec:impl-frag}, we provide a connection between macroscopic distinctness as formalized by our measure and fragility of entanglement to another system. We summarize and conclude in Sec.~\ref{sec:conclusion}.

\section{Macroscopically extractable information}
\label{sec:macr-inform-capac}

\subsection{Abstract definition}
\label{sec:definition}

Consider a set $\sS=\{ \sqrt{ p_\ell} \ket{\t{A}_\ell} \}_{\ell=1}^N$ of $N$ quantum states $\ket{A_\ell}$ and $N$ corresponding probabilities $p_\ell$. 
This set can be used to define a superposition state
\begin{equation}
\label{eq:superposition}
\ket{\Psi}_{\mathrm{S}} \propto \sum_{\ell=1}^{N} \sqrt{p_\ell} \ket{\t{A}_\ell},
\end{equation}
or of a ``micro-macro'' entangled state 
\begin{equation}
\label{eq:micro-Macro}
\ket{\Psi}_{\mathrm{mM}} = \sum_{\ell=1}^{N} \sqrt{p_\ell} \ket{\ell}_\t{m}\ket{\t{A}_\ell}_\t{M},
\end{equation}
between the system and some microscopic system with $N$ orthogonal states $\{ \ket{\ell} \}_{\ell=1}^N$ called ``the atom'' \footnote{This is in analogy to the radioactive atom in the thought experiment of Schrödinger. Since $N$ is large in general, note that the microscopic part is not literally expected to be a single atom with $N$ levels. Rather, we might think of it as a small system composed of $O(\log(N))$ particles.}. We wish to construct a meaningful definition for the size of such superposition states, based on some notion of generalized macroscopic distinctness of the superposed components $\sS$.

Following \cite{Sekatski_Size_2014}, we assume that we measure the macroscopic system with a measurement device that has a rather coarse-grained resolution $\Delta$ (i.e., ``low resolution'' means large $\Delta$). Let us consider a game in which Bob draws a random variable $\ell$ described by a probability distribution $p_\ell$ and sends the corresponding state $\ket{\t{A}_\ell}$ to Alice. She measures the received state with the detector (characterized by $\Delta$), and obtains some outcome $x$. The information that she collects on the random variable hold by Bob can be quantified by the mutual information (MI) of the probability distribution $p(\ell,x)$,
\begin{align}
\label{eq:mutual information}
I_\Delta(A:\ell)&=H(p_\ell) - \sum_x p(x) H(p(\ell|x)),
\end{align}
with the Shannon entropy $H(p_{\ell}) = -\sum_{\ell} p_{\ell} \log p_{\ell}$ \footnote{Our choice of the Shannon entropy here is not unique. One can easily think of contexts where other entropies would be more appropriate}.  Note that the MI can never exceed the Shanon entropy of the initial probability distribution $H(p_{\ell})$. Hence, the maximal MI for as set of $N$ orthogonal states with equal weights is given by $b_{\max}=\log_2(N)$.

The intuition inherited from the macroscopic distinctness of the cat's two states $\ket{\text{dead}}$ and $\ket{\text{alive}}$ tells us that a truly macroscopic superposition does not require technologically advanced detectors with high resolution in order to collapse the superposition to a single branch (or equivalently to learn the state of the atom in Eq.~(\ref{eq:micro-Macro})) \cite{Sekatski_Size_2014}. To quantify this intuition we define the effective size of $| \Psi \rangle_S $ or $| \Psi \rangle_{\mathrm{mM}} $ as the maximal $\Delta$ of the detector that still allows Alice to gain $b$ bits of information about the preparation of Bob 
\begin{equation}\label{eq:MIC}
\t{MIC}_b (\sS) = \max_\Delta\{ \Delta | I_\Delta(A:\ell)\geq b\},
\end{equation}
standing for the \emph{Macroscopicness of  Information Content} of the superposition.
The minimal information $b$ is a parameter of the proposed measure, whose role we discuss later.

\subsection{Model of a coarse-grained measurement}
\label{subsec:model imperf measurement}

Up to this point we were quite unspecific about the measurement device. Indeed, the definition above only assumes that there is a meaningful way to attribute a resolving parameter $\Delta$ to the measurement device (and to continuously vary this parameter). In general, the detector does not have to be uniquely characterized by $\Delta$, but can have additional knobs. In such a case, an additional optimization is necessary, as one is interested in the largest possible MI. However, we do not consider this more complicated situation in the following. As a first example, note that low resolution can come from inefficiencies modeled by a loss channel preceding an ideal measurement, in which case $\Delta$ is associated to the probability to not (or only partially) measure the system (see also \cite{Korsbakken_Measurement-based_2007,Sekatski_Difficult_2014}).

In the following, however, we will consider the von Neumann pointer model with weak coupling between system and pointer. Suppose one would like to measure system with the observable
\begin{equation}
A = \sum_{\ell} a_{\ell} \left| A_{\ell} \right\rangle\!\left\langle A_{\ell}\right|, 
\label{eq:observable}
\end{equation} 
which, for simplicity, is supposed to have non-degenerate discrete spectrum (if this is not the case replace the sum with an integral). For the formal definition of our measure the choice of $A$ is irrelevant. However, it does determine which states are considered to be macroscopically distinguishable. Typically, we choose operators $A$ with a classical limit such as collective spin operators for atomic ensembles or number of photons and quadrature operators for photonic state.

The measurement is done via a pointer $P$ (i.e., an auxiliary system), which first interacts with the system and is subsequently read out in a preferred basis. Consider a pointer system modeled by a particle on a one-dimensional line with the usual commutation relation for position and momentum  $[\hat{x},\hat{p}]=\ii$ (with $\hbar=1$). We assume the pointer's initial state to be 
\begin{equation}
\ket{\xi_\Delta}=\int \xi_\Delta(x) \ket{x} dx,
\label{eq:pointer}
\end{equation}
with $\Delta$ characterizing the width of the distribution $|\xi_\Delta(x)|^2$, and we choose a real valued function $\xi_\Delta(x)$. The system interacts with the pointer via the unitary $U=e^{-\ii \, A \otimes \hat{p}}$. Afterwards, the pointer is measured in the $x$-basis leaving the system in the state 
\begin{equation}
\rho_x = 
   \frac{K_x \rho K_x^\dag}{\tr K_x^\dag K_x \rho}
\end{equation}
with $K_x = \bra{x}U \ket{\xi_\Delta} = \xi_\Delta(x-A)$. On an abstract level, this protocol realizes a general measurement with POVM elements $K_x^2 = \xi_\Delta^2(x-A)$. Trivially, if the width $\Delta$ of the initial pointer state tends to zero, one recovers the usual ``strong'' projective measurement $\xi_\Delta^2(x-A) \to \delta(x-A)$. In contrast, the coupling becomes effectively weaker as $\Delta$ increases. The system is less disturbed by the measurement and, consequently, the measurement progressively loses resolution and becomes less informative. This is sometimes called a weak measurement.

In case one does not postselect on (or does not have access to) the measurement result $x$, the post-measurement state of the system --after tracing out the pointer-- reads
\begin{equation}\label{eq:dephasing}
\rho_\t{out}= \tr_P U \rho \otimes \prjct{\xi_\Delta} U^\dag = \int \mu(p) e^{- \ii p A} \rho \,e^{ \ii p A} dp,
\end{equation}
where $\mu(p)= |\!\bracket{p}{\xi_\delta} \!|^2$. In words, if the measurement outcome is ignored the effect of the weak measurement on the state is a dephasing channel generated by the observable $A$. Note that $\bracket{p}{\xi}$ and $\xi_\Delta(x)$ are connected via a Fourier transform, such that, in general, the weaker the measurement the lower is the strength of the induced dephasing.

To be more specific, we consider two examples for the pointer function $\xi_{\Delta}(x)$ in the following. In Sec.~\ref{sec:exampl-equally-spac}, we assume the distribution of the pointer to be square with a width $\Delta$, such that an outcome $x$ corresponds to a POVM element $E_\Delta(x- A)= \xi_\Delta^2(x- A)$ with
\begin{equation}\label{eq:flat pointer}
E_\Delta(x)=\begin{cases}
\frac{1}{\Delta}& |x| \leq \frac{\Delta}{2}\\
0 & \text{otherwise}.
\end{cases}
\end{equation}
Another important example is when $\xi_{\Delta}^2(x-A) = g_\Delta(x-A)$ is a Gaussian function with spread $\Delta$, that is,  \begin{equation}\label{eq:1}
g_\Delta(x) = \frac{1}{\sqrt{2\pi}\Delta} e^{-\frac{x^2}{2\Delta^2}}.
\end{equation}

\subsection{Example: Equally spaced peaks}
\label{sec:exampl-equally-spac}

We now illustrate our formalism with a simple example of the equally weighted superposition
\begin{equation}\label{eq:2}
\ket{\Psi}_{\mathrm{S}} = \frac{1}{\sqrt{k+1}}\sum_{\ell=0}^k \ket{\ell \frac{N}{k} }
\end{equation}
of $k+1$ equally spaced eigenstates $A\ket{l \frac{N}{k} } =\frac{\ell N}{k}\ket{\ell \frac{N}{k} }$, all contained in the interval $[0,N]$, and a square pointed $E_\Delta(x)$ of Eq.~\eqref{eq:flat pointer}. As the distribution is uniform one has $H(p_\ell)=\log_2(k+1)$. First, note that the probability to observe an outcome $x$ only depends on $k$ and the ratio $r=\frac{\Delta}{2 N}$. So we directly move to the scale-invariant problem with $k+1$ eigenstates $\sS=\{(1+k)^{-1/2}\ket{\frac{\ell}{k}}\}_{\ell=0}^k$ contained in the interval $[0,1]$, and the square pointer $E_{2r}(y)$ of width $2r=\frac{\Delta}{N}$.

\begin{figure}[!ht]
\centering
\includegraphics[width=0.98\columnwidth]{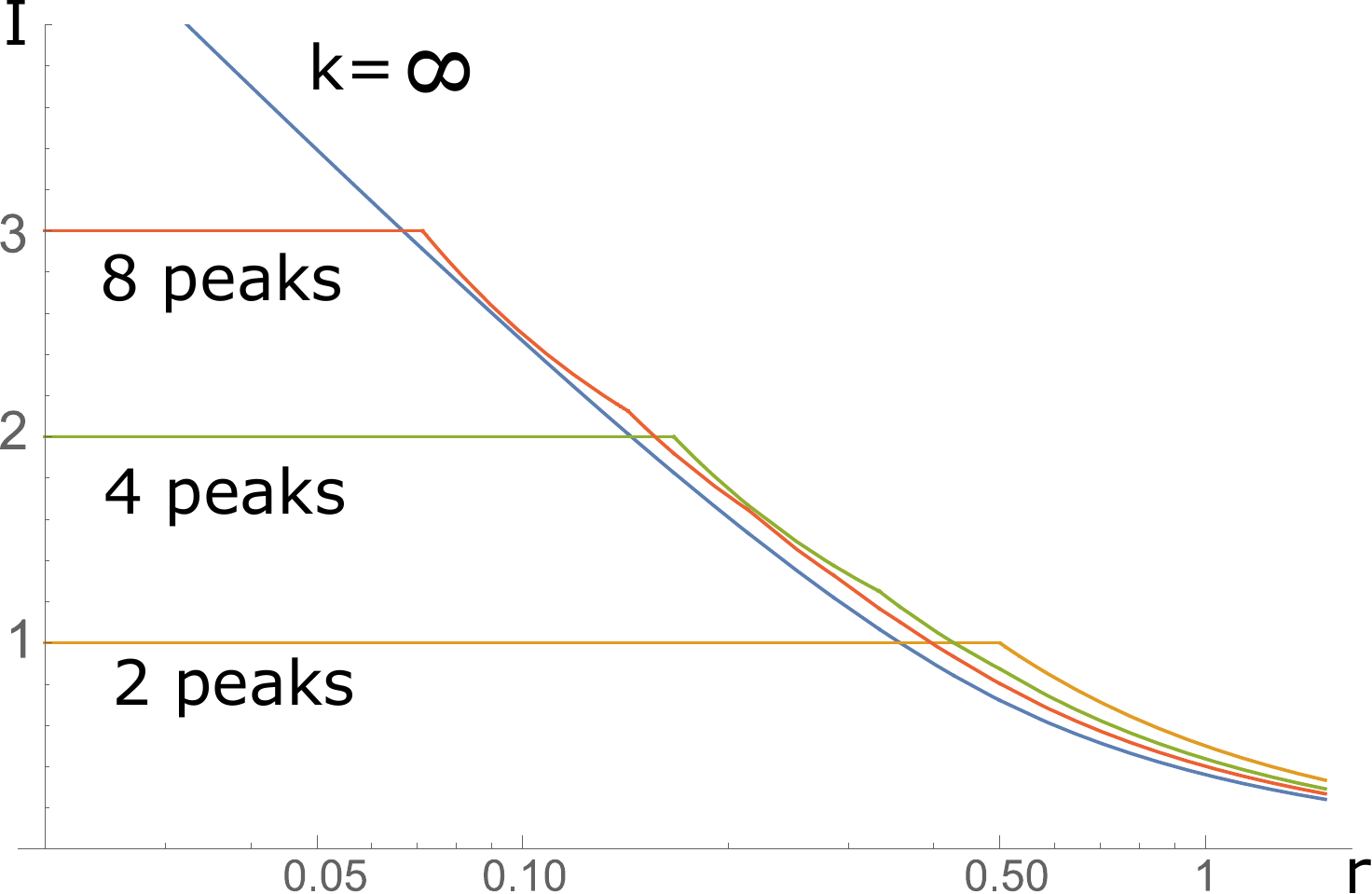}
\caption{Mutual information $I(A:\ell)$ as function of the ratio $ r= \frac{\Delta}{2N}$ for two, four, eight and infinite number of peaks. While large MI can only be reached with many peaks and small $r$, the two-peak state maximizes the MI within this state family for $r \geq 1/2$.  }
\label{fig:Npeaks}
\end{figure}

The calculation of the MI is mainly a combinatorial problem. Lengthy but straightforward arithmetics (see Appendix \ref{sec:deta-equally-spac}) gives, for $r\geq 1/2$,
\begin{equation}
P_n^\geq(r,k) =
\begin{cases}
 \frac{n}{k (k+1) r} & 1\leq n\leq k \\
 1-\frac{1}{2r} & n=k+1 \\
 0 & n>k+1 \\
\end{cases}
\end{equation}
and, for $r<1/2$ with $c=\floor{2r k}$,
\begin{equation}
P_n^<(r,k) =\begin{cases}
 \frac{2 n}{k (2 (k+1) r)} & 1\leq n\leq c-1 \\
 \frac{c (k-c) \left(\frac{c}{k}+\frac{1}{k}-2 r\right)+\frac{2 c}{k}}{2 (k+1) r} & n=c \\
 \frac{(c+1) (k-c+1) \left(2 r-\frac{c}{k}\right)}{2 (k+1) r} & n=c+1 \\
 0 & n>c+1.
\end{cases}
\end{equation}
This implies
\begin{equation}
I_\Delta(A:\ell)=
\begin{cases}
\frac{N}{\Delta}(\log_2(k+1) -\frac{2 \log_2(H!(k))}{ k (k+1)}) & \Delta \geq N\\
\sum_{n=1}^{c+1} P_n^<(\frac{\Delta}{2N},k) \log_2(n) & \Delta < N
\end{cases}
\end{equation}
with the hyperfactorial $H!(k)=\Pi_{n=1}^k n^n$. In Fig.~\ref{fig:Npeaks} we plot $I_\Delta$ for several numbers of peaks, as well as the limiting case $k\to \infty$. 

For $\Delta\geq N$ the maximal MI is obtained for two peaks and is given by $I_\Delta=\frac{N}{\Delta}$. Accordingly for $b\leq 1$ this is also the state that maximizes
\begin{equation}
\max_k \t{MIC}_{b\leq 1} = \frac{N}{b}
\end{equation}
For $\Delta<N$ things are more complicated, but numerical evidence shows that for $b=\log_2(k+1)$
\begin{equation}\label{eq:MIC many peaks}
\max_{k^{\prime}} \,\t{MIC}_{b=\log_2(k+1)} = \frac{N}{2^b-1} = \frac{N}{k}
\end{equation}
is maximized by the state with $k+1$ peaks.  Combining the two we find, for any $b$, the maximal size attained by state with eqally spaced peaks in the interval $[0,N]$
\begin{equation}\label{eq:MIC peaks}
\t{MIC}_{b,N}^\t{peaks} =
\begin{cases}
\frac{N}{b} & b\leq 1 \\
\frac{N}{2^b-1}& b>1
\end{cases}
\end{equation}

 To conclude this example let us remark that with the results above this family of states can be used for calibration of the measure $\t{MIC}_b$ for any state. Concretely, for any superposition state in addition to attributing a value $\t{MIC}_b$ for each $b$ one  says that the state under consideration is as macroscopic as $k+1$ equally spaced peaks in the interval $[0,N]$, for some $k$ and $N$ easily obtained from Eq.~\eqref{eq:MIC peaks} and Eq.~\eqref{eq:MIC many peaks}.

\subsection{Role of $b$ and calibration of the measure}

The proposed measure is parametrized by $b$, that is, the amount of extractable information in the protocol of Sec.~\ref{sec:definition} measured in bits. This might seem as a flaw of our approach, adding some arbitrariness to the definition. But this is not so, $b$ can be understood as the ``rank'' of the macroscopic superposition -- it counts the effective number of different components that are superposed. This is  an important characterization of the state that is independent and irreducible to its ``size''. For example, the state in the famous thought experiment of Schr\"odinger cat $\ket{\uparrow}\ket{\t{alive}} + \ket{\downarrow}\ket{\t{dead}}$ is undeniably a very large macroscopic superposition, still it is a superposition of only two components and can never yield more then one bit of information. Similarly, one can easily think of a microscopic state that is a superposition of many components yielding a large amount of information $b\gg1$, nevertheless it has a small size $\t{MIC}_{b=1}$ even for one bit.

It is then appealing to introduce an archetypal reference state for each value of $b$, which can be used for the calibration of the size measure. In view of the results above, this can be naturally done using the family of $k+1$-peaks states. Concretely, for any value of $b_k=\log_2(k+1)$ we can identify the state with $k+1$ equally spaced peaks in the interval $[0,N]$. Then for a general state $\ket{\Phi}_\t{S}$  and for each value $b_k$, in addition to attributing a value $\t{MIC}_{b_k}$, one can conclude that the state $\ket{\Psi}_\t{S}$ is as macroscopic as the state with $k+1$ peaks distributed on the interval of width 
\be
N_{b_k}(\ket{\Psi}_\t{S}) = \max_N \{N | \t{MIC}_{b_k,N}^\t{peaks}\leq \t{MIC}_{b_k}\},
\ee
using the result of Eq.~\eqref{eq:MIC peaks}. $N_{b_k}(\ket{\Psi}_\t{S}) $ can be interpreted as a calibration of the size measure.

\subsection{Connection to the variance}
\label{sec:infl-b-conn}

The variance of a state $V(\ket{\Psi},A) = \bra{\Psi}A^2\ket{\Psi} -\bra{\Psi}A\ket{\Psi}^2$ is a natural measure of how large is the spread of a state in the eigenbasis of $A$. So it is natural to study the relation of our measure to the variance. For this, we consider Gaussian pointers, Eq.~(\ref{eq:1}), for which the MI can be expressed as
\begin{align}\label{eqn:variance_def}
	I_\Delta(A : \ell) & = H(p(x)) - \sum_\ell p_\ell H(p(x|\ell)) \\
	& = H(p(x)) - \frac{1}{2} \log(2 \pi e \Delta^2),
\end{align}
since $p(x|\ell) = g_\Delta(x-a_\ell)$. In Appendix \ref{app:I_weak_limit}, we prove that MI is always upper bounded by the variance
\begin{equation} \label{eqn:variance_ineq}
	I_\Delta(A:\ell) \leq \frac{V(\ket{\Psi},A)}{(2 \ln 2) \Delta ^2} \; \quad \forall \Delta.
      \end{equation} 
One might wonder if there also exists a lower bound involving the variance. However, with the following example it is easy to see that no such bound can exist. For an appropriate choice of parameters $p$ and $N$ the superposition state $\sqrt{p} \ket{0}+\sqrt{1-p} \ket{N}$ can have an arbitrarily low MI and an arbitrarily large variance. Consequently, the two are inequivalent and the requirement for a large MI is strictly more restrictive than for a large variance.     
Nevertheless, the inequality \eqref{eqn:variance_ineq} becomes tight when $\Delta$ is sufficiently large 
\begin{equation} \label{eqn:variance_approx}
	I_\Delta(A:\ell) \approx \frac{V(\ket{\Psi},A)}{(2 \ln 2) \Delta ^2}.
      \end{equation}

This shows that, for a weak Gaussian measurement and for pure states, our measure is connected to earlier proposals \cite{Bjork_Size_2004,Shimizu_Detection_2005,Lee_Quantification_2011,Frowis_Measures_2012} where the variance $V(\ket{\Psi},A)$ plays a role to measure the macroscopic distinctness. Equation (\ref{eqn:variance_approx}) is further useful to evaluate our measure for small $b$.

\section{Mixed states and convex roof}
\label{sec:mixed-states-convex}

In practice, quantum states $\rho$ are mixed. On the conceptual level, one can treat the \emph{macroscopicness} 
and the \emph{quantumness} of $\rho$ as two independent aspects. The mixedness of a state $\rho$ can then be attributed to the decay of its quantumness, while its maroscopicness, stemming from $\sS$, is left unchanged. Nevertheless this is not satisfactory in our case. First, we would like the MIC measure to be a single quantity that encompasses both the macroscopicness and the quantumness of the state. Second, a mixed state $\rho= \sum q_i \prjct{\Psi_i}$ admits infinitely many ensemble decompositions which can yield different average MIC, since different elements $\ket{\Psi_i}$  correspond to different $\sS_i$ and do not necessarily have the same size. 

To get a MIC defined on all states $\rho$ and non-increasing on average under mixing one uses the convex-roof extension 
\begin{equation}\label{eq:convex roof}
\widehat{\t{MIC}}_b (\rho) = \min_{\sum_k q_k \prjct{\Psi_k}=\rho} \sum_k q_k \t{MIC}_b (\ket{\Psi_k}).
\end{equation} 
In words one finds the ensemble partition of $\sum q_i \prjct{\Psi_i}=\rho$ that has the least average size, and defines this value as the size of $\rho$.
This is by construction non-increasing under mixing, given any measure defined on pure states. Note that, despite the uncountable number of pure-state decompositions of $\rho$, the number of pure states in an extremal ensemble is limited to $d^2$, where $d$ is the rank of $\rho$.  They also form a closed manifold, as there is a one to one mapping between decompositions of $\rho$ and partitions of identity, or POVMs, see appendix \ref{sec:finite-number-pure}).

As an example, we consider quantum states lying in the span of two eigenstate of the observable $\ket{0}$ and $\ket{N}$ as in Sec.~\ref{sec:exampl-equally-spac}. The most general state of this form reads 
\begin{equation}\label{eq:state2peaks}
\rho= \frac{1}{2}\left(\begin{array}{cc}
1+z_\rho & x_\rho-iy_{\rho} \\
x_\rho+iy_{\rho} &1-z_\rho 
\end{array}\right),
\end{equation}
expressed in the basis $\{\ket{0},\ket{N}\}$ in the superposition scenario, or $\{\ket{0,0}_{\mathrm{mM}},\ket{N,N}_{\mathrm{mM}}\}$ in the micro-macro entanglement scenario. To shorten the notation we define ${\bf r} = (x_\rho\, y_\rho\, z_\rho)$. As the size is invariant under rotation of the state around the z-axis, we assumed $y_\rho=0$ in Eq.~\eqref{eq:state2peaks}.

For pure states (i.e., ${\bf r}^2=1$) and a square pointer as defined in Eq.~\eqref{eq:flat pointer}, the MI Eq.~\eqref{eq:mutual information} can be easily computed
\begin{equation}
\tilde I_L(x)= \tilde I_0(x)\, \min(\frac{N}{\Delta},1)
\end{equation}
with a convex function
\begin{equation}
\tilde I_0(x) = 1 -\log(|x|) +\frac{\sqrt{1-x^2}}{2}\log\left(\frac{1-\sqrt{1-x^2}}{1+\sqrt{1-x^2}}\right)
\end{equation}
Consequently, for pure states the size is given by 
\begin{equation}\label{eq:size x}
\t{MIC}_b(x) =N  \begin{cases} \frac{\tilde I_0(x)}{b} & I_0(x)>b \\
0 & \text{otherwise},
\end{cases} 
\end{equation}
where the spread of the state $N$ appears as a factor (as only the relative size of the spread to the pointer width is relevant).

\begin{figure}[!ht]
\centering
\includegraphics[width=0.98\columnwidth]{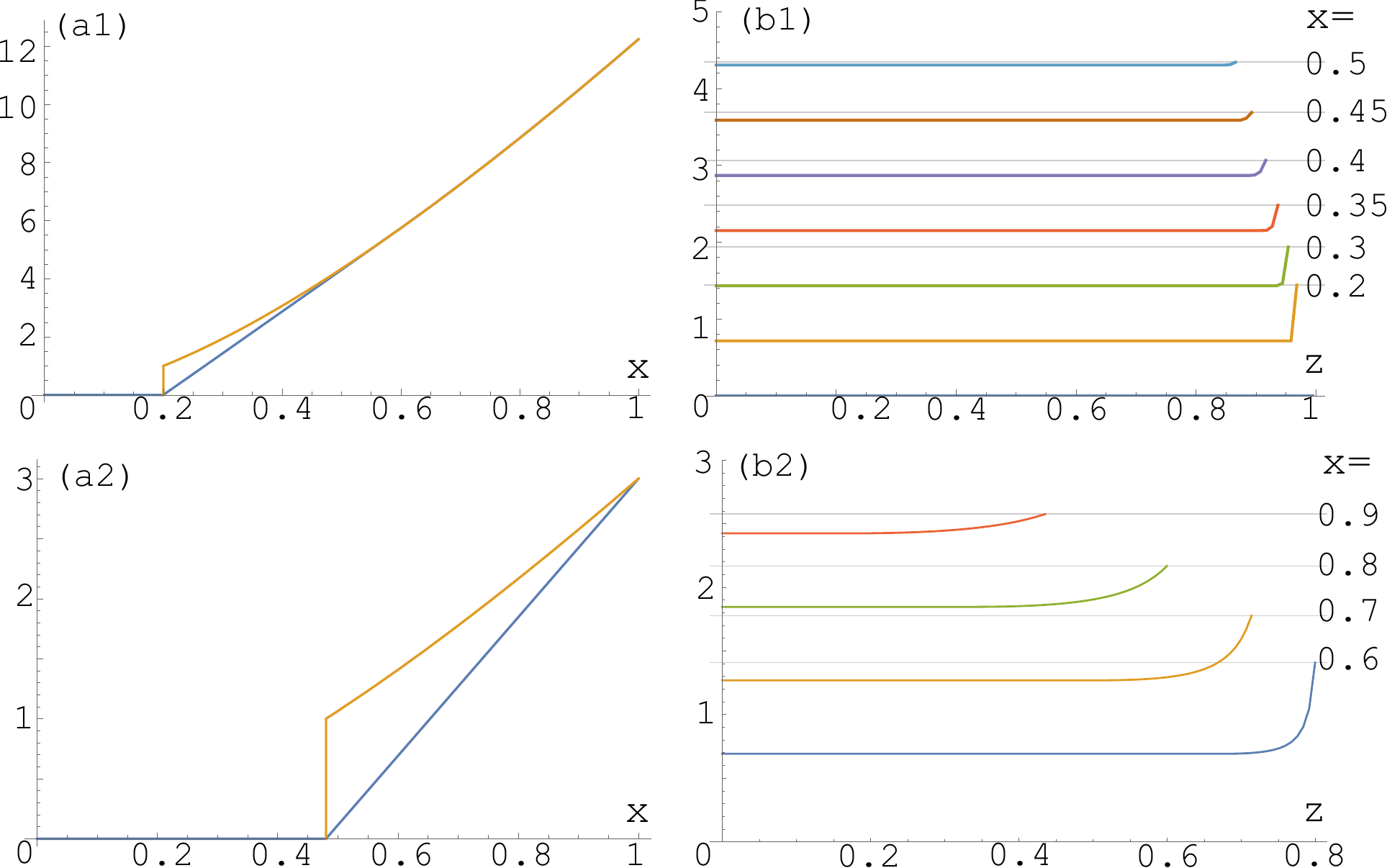}
\caption{(a1-a2) Rescaled size $\widehat{ \t{mic}}(x_\rho, z_\rho)$ of the states as function of $x_\rho$ for $b = 0.082$(a1) and $b = 1/3$(a2). The top curves corresponds to pure states with $z_\rho=\sqrt{1-x_\rho^2}$, and the bottom curve corresponds to states with $z_\rho=0$. (b1-b2) Rescaled size $\widehat{ \t{mic}}(x_\rho, z_\rho)$ of the state as function of $z_\rho$ for $b = 0.082$(b1) and $b = 1/3$(a2). Different curves correspond to different constant values of $x_\rho$ given on the right.}
\label{fig:Roof}
\end{figure}

Given this expression  one can compute the average size for any ensemble partition of $\rho$. It remains then to find the optimal ensemble. Though this step can be done analytically, it turns out to be quite tedious (see Appendix \ref{app:roof}). The final result is a function 
\begin{equation}
\widehat{ \t{MIC}}_b(x_\rho, z_\rho,N) =N\, \widehat{ \t{mic}}_b(x_\rho, z_\rho),
\end{equation}
 that linearly depends on $N$, but has a complicated dependence of the state $\rho$.

 In Figure~\ref{fig:Roof} we plot the rescaled size $\widehat{ \t{mic}}_b(x_\rho, z_\rho)$  for two values of $b=0.082$ and $b=1/3$. These value are chosen to allow a comparison with the example of \cite{Sekatski_Size_2014}, where equally-weighted pure superpositions $\frac{1}{\sqrt{2}}\left(\ket{0}+ \ket{N}\right)$ are characterized by the probability $P_c$ to correctly distinguish between the two branches ($| 0 \rangle \equiv \left| A \right\rangle $ and $| N \rangle \equiv \left| D \right\rangle $ in \cite{Sekatski_Size_2014}; with the choice $P_c=2/3$ in the example). The value of $P_c$ does not uniquely determine the MI obtained by the measurement. Indeed, as entropy is concave given the two distributions $p(x|0)$ and $p(x|N)$, with the fixed guessing probability 
\begin{equation}
 P_c \equiv \frac{1}{2}\int \max\big({p(0|x),p(N|x)}\big) dx=2/3
\end{equation}
 the MI can range from $0.082$ bits for the extremal case, where for each outcome $x$ either $p(0|x)=2/3$ and $p(N|x)=1/3$ or vice versa, to $1/3$ bits for the other extreme, where for with probability $1/3$ one obtains an outcome for which either $p(0|x)=0$ or $p(N|x)=0$ and $p(0|x)=p(N|x)=1/2$ for all the others others.

\section{Alternatives and comparison}
\label{sec:altern-comp}

The convex-roof extension for mixed states, Eq.~(\ref{eq:MIC}), is conceptually straightforward. However, the convex roof of an inverted function is generally difficult to handle in proofs or in calculations of specific examples. In this section, we present alternative formulations and compare them to recent contributions in the literature. We present two variants. For the first one, we start with the original idea using the MI, but do the convex-roof extension of the MI instead of MIC. The second alternative uses a slightly different motivation to directly measure the nonclassical part of the macroscopically extractable information. We find a formulation that turns out to be equivalent to the so-called basis-dependent quantum discord.

\subsection{Direct convex-roof extension of the MI}
\label{sec:direct-convex-roof}

Instead of the convex-roof extension of MIC we consider the direct convex roof of the MI
\begin{equation}
	I_\Delta(A:\ell)_\rho = \min_{q_k,\ket{\Psi_k}} \sum_k q_k I_\Delta(A:\ell)_{\ket{\Psi_k}}
\end{equation}
and define a slightly different version of the MIC, namely 
\be
\widehat{\t{MIC}}'_b(\rho) = \max \{\Delta | I_\Delta(A:\ell)_\rho \geq b\}.
\ee
For the example of Eq.~\eqref{eq:state2peaks} this definition gives the size of a pure state $\widehat{\t{MIC}}'_b(\rho) =\t{MIC}_b(x_\rho)$ with the same $x_\rho$ in Eq.~\eqref{eq:size x}, as it follows from the convexity of $\tilde I_0(x)$. So it has the advantage to be more straightforward to compute. In addition this alternative definition allows us find the following connection.

In \cite{Yadin_General_2016}, a set of criteria were proposed for quantities that aim to capture the macroscopic coherence of a state. These are in the same spirit as the criteria for good entanglement measures. The most important ones say that a valid measure should (C1) vanish if and only if a state is an ``incoherent'' mixture of the form $\sum_\ell p_\ell \prjct{A_\ell}$; (C2) not increase under any ``covariant'' operation. An operation is covariant when it commutes with transformations of the form $e^{-i t A}$ -- this set captures all the possible operations which cannot create a superposition of the $\ket{A_\ell}$ and which respect the ``scale'' $|a_i - a_j|$ of a superposition $\ket{A_i}+\ket{A_j}$. (C2) can be broken down into two versions, (C2a) for deterministic processes, and (C2b) for stochastic processes under which the measure cannot increase on average. In addition, one can demand that a measure be (C3) convex (i.e., non-increasing under mixing) and (C4) increasing with respect to the scale $|a_i-a_j|$.

We show in Appendix \ref{app:I_criteria} that this extended $I_\Delta(A:\ell)_{\rho}$, assuming a Gaussian pointer, satisfies all criteria (C1-4). In addition, $\widehat{\t{MIC}}'_b$ satisfies (C2a), (C4) and a modified version of (C1), namely
\begin{equation}
	\widehat{\t{MIC}}'_b = 0 \Leftrightarrow I_\Delta(A:\ell) \leq b \; \forall \Delta.
\end{equation}
In other words, $\widehat{\t{MIC}}$ is well-behaved in the sense that it vanishes for states that are close to incoherent mixtures, cannot increase under covariant operations, and is increasing with the scale of a superposition.

It is also worth noting that the relations between the MI and the variance for pure states of Eq.~\ref{eqn:variance_ineq}are directly generalized to mixed states via the convex roof of MI
\begin{align}
I_\Delta(A:\ell)_\rho &\leq \frac{\mc{F}(\rho,A)}{(8 \ln 2)\Delta^2}\\
I_\Delta(A:\ell)_\rho & \approx \frac{\mc{F}(\rho,A)}{(8 \ln 2)\Delta^2} \quad \t{for} \quad b\to 0.
\end{align}
Where  the quantum Fisher information $\mc{F}(\rho,A)$ \cite{Braunstein_Statistical_1994} of the state $\rho$ with respect to the operator $A$ is known to equal to four times the convex roof of the variance \cite{Yu_Quantum_2013}. It follows that $\widehat{\t{MIC}}'_b$ satisfies
\begin{align}
	\widehat{\t{MIC}}'_b(\rho) & \leq \sqrt{\frac{\mc{F}(\rho,A)}{(8 \ln 2) b}} \quad \t{and} \\
	\widehat{\t{MIC}}'_b(\rho) & \approx \sqrt{\frac{\mc{F}(\rho,A)}{(8 \ln 2) b}} \quad \t{for} \quad b\to 0.
\end{align}
This allows one to obtain upper-bounds on the size. Moreover, since the quantum Fisher information is known to satisfy all criteria (C1-4), we conclude that $\widehat{\t{MIC}}'_b$ fulfills them in the limit of $b\rightarrow 0$. Note that the quantum Fisher information plays a central role in one of the general proposals for macroscopic distinctness \cite{Frowis_Measures_2012}, so this measure is in some sense contained in the presented family as a limiting case.

In particular, the insights we have about the quantum Fisher information can be used to apply our measure for small $b$ to real experimental data \cite{Frowis_Lower_2017}.
\subsection{Alternative measure using quantum correlations}
\label{sec:altern-meas-using}

In this section, we build up an alternative measure which is conceptually similar but has a slightly different motivation. As argued earlier, the distinguishability of a set of states under a noisy measurement can be captured by the mutual information between Bob, who prepares the ensemble of states, and Alice, who reads out the measurement device. Put differently, when a measurement device interacts with a system in the superposition state $\ket{\Psi}_{\mathrm{S}} = \sum_\ell \sqrt{p_\ell} \ket{A_\ell}$, the correlations $I(A:\ell)$ between the macroscopic system $M$ and measurement device $P$  are related to how well the device discriminates the branches of the superposition. However, $I(A:\ell)$ can be non-zero even when the system is initially in an incoherent mixture $\sum_\ell p_\ell \prjct{A_\ell}$. This issue can be avoided by using the convex roof constructions for mixed states. This is conceptually appealing, but comes at a price of make things hard to compute, as illustrated with the example of Section~\ref{sec:exampl-equally-spac}. But is there a more direct way to avoid this issue that retains the nice physical intuition behind the MI. 

Here we introduce the quantity $C_\Delta(\rho, A)$ that is related to $I_\Delta(A:\ell)$ in spirit, but can be directly applied to mixed states. We start by introducing it in two different ways, and then show that they are equivalent. Let the system start in an arbitrary state $\rho$, and the measurement device in the initial pointer state $\ket{\xi_\Delta}$. We call 
\be
\rho' = U \rho \otimes \prjct{\xi_\Delta} U^\dag
\ee 
the overall state after the interaction $U = e^{-i A \otimes \hat{p}}$,  $\rho'_M$ the post-measurement state of the system, given in Eq.~\eqref{eq:dephasing}, and $\rho'_P$ the final state for the pointer. Using the Von Neumann entropy $S(\rho) = - \mathrm{Tr} \rho \log \rho$, define
\begin{equation}
	C_\Delta(\rho,A) := S(\rho'_M) - S(\rho)
\end{equation}
as the entropy difference between the post-measurement state $\rho_M'$ , given in Eq.~\eqref{eq:dephasing}, and the initial state $\rho$. Intuitively, the entropy increase in the system can only come from its correlations to the pointer created by the interaction. Hence, $C_\Delta(\rho,A)$ captures how much information is potentially available to the pointer about the system\footnote{In contrast to $I(A:\ell)$, this quantum information is not necessarily completely extractable, due to discord. Nor is it necessarily information about $\ell$, as  the states $\{ \ket{A_\ell}\}$ is only one particular basis choice for the system.} Note that this quantity avoids all problems associated with mixed states. In particular, the  system state with no coherence
\be
\mathcal{G}(\rho) :=\sum_\ell \prjct{A_\ell} \rho \prjct{A_\ell} =\sum_\ell p_\ell \prjct{A_\ell}
\ee 
is not affected by the interaction with pointer, implying $C_\Delta(\mathcal{G}(\rho),A) =0$.

An alternative definition can be given via the quantum mutual information (QMI) $\mathcal{I}(P:M)_{\rho'} \mathrel{\mathop:}= S(\rho_P') + S(\rho_M') - S(\rho')$ between the system and the pointer after the interaction. As we show in the next paragraph,
\be\label{eq:C disc}
C_\Delta(\rho,A) := \mathcal{I}(P:M)_{\rho'} - \mathcal{I}(P:M)_{\mathcal{G}(\rho)'}.
\ee
is given by the QMI for the initial state $\rho$ minus the QMI for its incoherent version $\mathcal{G}(\rho)$. Here, the issue of mixed states in resolved even more explicitly as the incoherent contribution to the QMI is simply subtracted. In fact, the definition \eqref{eq:C disc} also makes it clear that $C_\Delta(\rho,A)$ corresponds to the (fixed-basis) quantum discord of the final state \cite{Ollivier_Quantum_2001}. As the quantity
\be
\mathcal{I}(P:M)_{\mathcal{G}(\rho')} =S(P|M)=\sum_\ell p_\ell S(\rho_{P| \ell}),
\ee 
with $\rho_{P| \ell} =\frac{\tr_M \rho' \prjct{A_\ell}}{\tr \rho' \prjct{A_\ell}}$ and $\tr \rho' \prjct{A_\ell}=p_\ell$,  is equal to the conditional entropy of the pointer on the system measured in the eigenstates of $A$ (note that the measurement commutes with the interaction). So since the initial state of the pointer is pure $\mathcal{I}(P:M)_{\mathcal{G}(\rho)'}=\mathcal{J}(P|M)_{\rho'}$ gives the QMI between the system and the pointer, available upon the measurement of $A$. Note that the usual approach is to maximize the classical correlations over all possible measurements on $M$; but since we have a fixed observable $A$ of interest here, it is natural to fix the measurement.

Let us now show that the two definitions are equivalent. Since the interaction is unitary, we have $S(\rho') = S(\rho \otimes \prjct{\xi_\Delta}) = S(\rho)$, thus $\mathcal{I}(P:M)_{\rho'} = S(\rho'_M) + S(\rho'_P) - S(\rho')$. For the classical correlations, we write $\sigma=\mathcal{G}(\rho)$ and $\sigma' = \sum_\ell p_\ell \prjct{A_\ell} \otimes  e^{-i a_\ell p} \prjct{\xi_\Delta} e^{i a_\ell p}$. It follows that
\begin{align}
\mathcal{J}(P|M)_{\rho'} & = S(\sigma'_M) + S(\sigma'_P) - S(\sigma') \nonumber \\
& = H(p_\ell) + S(\rho'_P) - H(p_\ell) \nonumber \\
& = S(\rho'_P).
\end{align}
Therefore $C_\Delta(\rho,A) = S(\rho'_M) + S(\rho'_P) - S(\rho) - S(\rho'_P) = S(\rho'_M) - S(\rho)$.

So now we have the quantity $C_\Delta(\rho,A)$ in place of $I_\Delta(A:\ell)$. Note that, in the case of pure states, there is an inequality
\begin{equation}
	C_\Delta(\ket{\Psi},A) \geq I_\Delta(A:\ell).
\end{equation}
This is because a pure post-interaction state of $M$ and $P$ has $\mathcal{I}(P:M)_{\rho'} = 2 \mathcal{J}(P|M)_{\rho'} = 2S(\rho'_M)$, so $C_\Delta(\ket{\Psi},A) = J(P|M)$. Then observe that $I_\Delta(A:\ell)$ measures the correlations of $\sigma'$ with respect to a measurement of the pointer observable -- which cannot exceed the mutual information $\mathcal{I}(P:M)_{\mathcal{G}(\rho)'}$.

The rest of this section is devoted to examining the properties of $C_\Delta(\rho,A)$. We assume a Gaussian pointer from now on. As before, we can define another version of MIC:
\begin{equation}
	\widetilde{\t{MIC}}_b(\rho) = \max \{ \Delta | C_\Delta(\rho,A) \geq b \}.
\end{equation}

Just as before, we find that $C_\Delta$ satisfies all the coherence measure criteria (C1-4) -- see Appendix \ref{app:C_criteria} for the proof. Again, $\widetilde{\t{MIC}}_b$ satisfies (C2a), (C4) and a modified version of (C1),
\begin{equation}
	\widetilde{\t{MIC}}_b(\rho) = 0 \Leftrightarrow C_\Delta(\rho,A) \leq b \; \forall \Delta.
\end{equation}

Let us look at the behaviour of $C_\Delta$ in the limit of a weak Gaussian measurement, where $\Delta$ is large. We find that, for pure states, to leading order,
\begin{equation} \label{eqn:variance_approx2}
	C_\Delta(\ket{\Psi},A) \approx \frac{h(\Delta^{-2})}{4} V(\ket{\Psi},A),
\end{equation}
where $h(t) = -t \log t$. This contrasts a little with Eq.~(\ref{eqn:variance_approx}) for $I_\Delta(A:\ell)$, where the leading order was $\Delta^{-2}$. See Appendix \ref{app:C_weak_limit} for the proof.

Recently, another measure for macroscopic distinctness has been proposed \cite{Kwon_Disturbance-Based_2016}. Although formulated differently, it is closely related to the quantity $C_\Delta(\rho,A)$. In particular, both measures  fulfill the proposed set of criteria for macroscopic coherence \cite{Yadin_General_2016}. As noted in \cite{Kwon_Disturbance-Based_2016}, small $\Delta$ leads to the counter-intuitive result that the measure assigns a larger value to some product states than to a superposition of two extremal eigenstates of $A$. This reveals the role of $\Delta$ in these measures as a characteristic scale. Using the measure tells us how much quantum coherence the state of interest provides on this scale.
In our framework it is $\widetilde{\t{MIC}}_b(\rho)$, rather then $C_\Delta(\rho,A)$, that is quantifying the macroscopic quantumness of the state. But along the same line $\widetilde{\t{MIC}}_b(\rho)$ can be understood as the maximal scale at which the state provides the desired amount of coherence.

\section{Implications on Fragility}
\label{sec:impl-frag}

In this section we consider the micro-macro entangled state 
\begin{equation}
\ket{\Psi}_{\mathrm{mM}} = \sum_{\ell=1}^{N} \sqrt{p_\ell} \ket{\ell}_\t{m}\ket{\t{A}_\ell}_\t{M}. \nonumber
\end{equation}
Under the assumption $\bracket{A_j}{A_k} = \delta_{jk}$  Eq.~\eqref{eq:micro-Macro} gives the Schmidt decomposition of $\ket{\Psi}_\mathrm{\mathrm{mM}}$. For the particular case $\frac{1}{\sqrt{2}}\left(\ket{0}_\t{m} \ket{A}_\t{M}+ \ket{1}_\t{m} \ket{D}_\t{M}\right) $ we know that the micro-macro entanglement is more fragile for a larger size of the macroscopic part of the state using the framework of \cite{Sekatski_Size_2014,Sekatski_Difficult_2014}. Here we will show that a similar relation between the size of the state as defined in Eq.~\eqref{eq:MIC} and the fragility of entanglement under certain type of noise persists in the general case. The intuition behind is rather simple: If the noise channel can be interpreted as an imprecise measurement of the system by the environment, then the size of the state relates to the amount of information extractable by the environment. The decay of entanglement through the channel is related to the information obtained by the environment. Since at least the mathematical modeling of the environment and a measurement pointer is similar, we denote the environment as $P$ as well.

\subsection{Entanglement of formation}
\label{sec:entangl-form}

Concretely, we consider the entanglement of formation $E_F(\rho_{AB})$. $E_F$ is an entanglement measure \cite{Horodecki_Quantum_2009} on bipartite states, defined as the convex roof of the entropy of entanglement
\begin{equation}
E_F(\rho_{AB}) = \min_{\sqrt{q_k}\ket{\Psi_k}} \sum_k q_k \, S(\rho_k^B),
\end{equation}
where the entropy of entanglement is by definition an entanglement measure on bipartite pure states given by the Von Neumann entropy $S(\rho_k^B)=S(\rho_k^A)$ of the partial states $\rho_k^B = \tr_A \prjct{\Psi_k}$.  Because we assume all the branches of $\ket{\Psi}_{\mathrm{mM}}$ to be orthogonal, its entanglement of formation reads
\begin{equation}
E_F\left(\ket{\Psi}_{\mathrm{mM}}\right) = H(p_\ell).
\end{equation}
Note that the entanglement in the state is invariant under local unitary transformation, so its amount is independent of the spread of the state in the spectrum of A and of the macroscopicness of the state.

\begin{figure}[!ht]
\centering
\includegraphics[width=0.7\columnwidth]{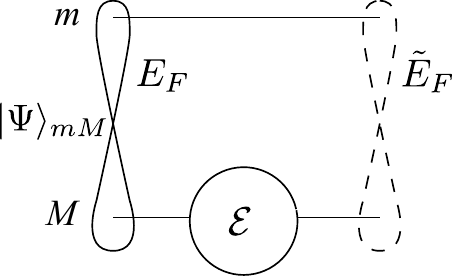}
\caption{Noise channel $\cE_X^{\Delta}$ acting on the macroscopic part of the state $\ket{\Psi}_{\mathrm{mM}}$.}
\label{fig:Fragility}
\end{figure}

\subsection{Noise as measurement by environment}
\label{sec:noise-as-measurement}

For any Kraus representation ${\bf K}$ of a channel $\cE$ 
\begin{align}
\cE(\rho) &= \sum_x K_x \rho K_x^\dag \nonumber \\&= \tr (\sum_x{K_x \otimes \coh{x}{0}_P})\rho \otimes \prjct{0}_P (\sum_x{K_x^\dag \otimes \coh{0}{x}_P})
\end{align}
can be interpreted as a measurement of the system by the environment, described by the POVM elements induced by the Kraus operators $\{E_x= K_x^\dag K_x \}$ (in the expression above the sum is replaced by an integral if a Kraus representation is continuous). For simplicity, the channel is supposed to act only on the macroscopic part (see Fig.~\ref{fig:Fragility}). Hence, the output of the channel
\begin{equation}
\cE(\rho)= \sum_x p(x) \rho_x 
\end{equation} is a mixture of states $\rho_x =\frac{K_x \rho K_x^\dag}{p(x)}$ with $p(x)= \tr E_x \rho$ conditional on the environment observing outcome $x$. Note that the POVM elements do not uniquely specify the channel, as the same element $E_x$ can correspond to physically different Kraus operators $U K_x=\sqrt{E_x}$.

Now consider the action of the channel on the state $\rho =\prjct{\Psi}_{\mathrm{mM}}$, for which all the conditional states 
\begin{equation}
\rho_x=\prjct{\Psi_x}\quad \t{with} \quad \ket{\Psi_x}= \frac{K_x \ket{\Psi}_{\mathrm{mM}}}{\sqrt{p(x)}}
\end{equation}
 are also pure. Similarly to Eq.~\eqref{eq:mutual information} we define the MI between the microscopic system and the measurement (that arises from the Kraus representation $\bf K$ of the channel) carried by the environment $I_{\cE,\bf K}(P:\ell)$. One has
\begin{equation}\label{eq:mean Shanon}
\int  p(x) H(p(\ell|x)) dx =  H(p_\ell) - I_{\cE,\bf K}(P:\ell),
\end{equation}
where $p(\ell|x) =\tr \prjct{\ell}_\t{m} \rho_x$. Note that it does not matter whether the projection $\{\prjct{\ell}\}$ on the atom's side or the measurement $\{ E_x\}$ by the environment is performed first. The Shanon entropy of the distribution $p(\ell|x)$ upper-bounds the Von Neumann entropy of the partial states $\rho_x^\t{M} = \tr_\t{m} \prjct{\Psi_x}$ and $\rho_x^\t{m} = \tr_\t{M} \prjct{\Psi_x}$ \footnote{The most direct way to argue is that $H(p(\ell|x))$ corresponds to the Von Neumann entropy of the partial state $\rho_x^\t{m}$ after it undergoes a projection map in the basis $\{\ket{\ell}\}$, while the entropy is non-increasing under physical maps.}
\begin{equation}
S(\rho_x^\t{M})=S(\rho_x^\t{m}) \leq H(p(\ell|x)).
\end{equation}
This inequality allows one to obtain a bound on the average partial entropy of the post-channel state
\begin{equation}\label{eq:mean Von Neumann}
\sum_x p(x) S(\rho_x^\t{M}) dx \leq  H(p_\ell) - I_{\cE,\bf K}(P:\ell).
\end{equation}

The left hand side is the average entropy of entanglement of the state $\cE(\ket{\Psi}_{\mathrm{mM}})$ that correspond to its pure-state partition provided by the Kraus representation $\bf K$. Consequently, by definition of the entanglement of formation one has 
\begin{equation}
E_F\Big(\cE(\ket{\Psi}_{\mathrm{mM}}) \Big) \leq E_F\left(\ket{\Psi}_{\mathrm{mM}}\right) - I_{\cE,\bf K}(E:\ell).
\end{equation}
In words, the decay of entanglement of formation through a channel is lower or equal than the MI obtained by the environment via the measurement induced by \emph{any} Kraus representation by the channel. 

\subsection{Examples}
\label{sec:examples}

  (i) Dephasing generated by the observable $A$
\begin{equation}
\cE_X^\delta (\rho) = \int \mu(\lambda) e^{-\ii\lambda A} \rho\,  e^{\ii\lambda A}  d\lambda,
\end{equation} 
of strength $\delta$ given by the with of the distribution $\mu(\lambda)$. As already mentionened in \eqref{eq:dephasing}, this noise corresponds to a coarse-grained measurement 
of $A$ by the environment. And the noise distribution $\mu(p)= |\!\bracket{p}{\xi}\!|^2$ is related to the resolution of the measurement $\xi_\Delta^2(x)=|\!\bracket{x}{\xi}\!|^2$ by a Fourrier transform implying $\delta \sim \frac{1}{\Delta}$. This shows that the quantity $I_\Delta(A:\ell)$ yields a lower bound on the decrease of entanglement in the state after the action of the channel $\cE_X^\Delta$. Similarly, $1/\t{MIC}_b\big(\ket{\Psi}_{\mathrm{mM}}\big)$ gives an upper bound on the amount of noise $\delta \sim \frac{1}{\Delta}$ that leaves $E_F\big(\ket{\Psi}_{\mathrm{mM}}\big)-b$ bits of entanglement in the system.

In the case of a channel $\mc{E}$ describing weak Gaussian noise from the environment, Eq.~(\ref{eqn:variance_approx}) shows that
\begin{equation}\label{eq:4}
	E_F \big(\mc{E}(\ket{\Psi}_{\mathrm{mM}}\big) \lesssim E_F(\ket{\Psi}_{\mathrm{mM}}) - \frac{V(\ket{\Psi}_{\mathrm{mM}},A)}{(2 \ln 2) \Delta^2}.
      \end{equation}

It also turns out that $C_\Delta$ lets us say something about the degradation of quantum correlations between $m$ and $M$. Note that $\mc{E}(\ket{\Psi}_{\mathrm{mM}})$ has the structure of a ``maximally correlated state'', which is generally written as $\sum_{i,j} \rho_{ij} \coh{i}{j} \otimes \coh{i}{j}$. It is known that the entanglement of a maximally correlated state is often the same as the coherence of the corresponding single-system state $\sum_{i,j} \rho_{ij} \coh{i}{j}$ \cite{Winter_Operational_2016,Chitambar_Assisted_2016,Streltsov_Entanglement_2016}. For example, this is true for the distillable entanglement $E_D$ \cite{Horodecki_Quantum_2009} and the relative entropy of coherence $C_R$ \cite{Baumgratz_Quantifying_2014}, which can be written as
\begin{equation}
	C_R(\rho) = S(\rho || \mathcal{G}(\rho)),
\end{equation}
that is, the relative entropy between a state and its fully dephased version. It also has the simple expression $C_R(\rho) = S(\mathcal{G}(\rho)) - S(\rho)$.

The channel $\mc{E}$ leaves the fully dephased part of a state unchanged. Therefore we simply have
\begin{align}
	E_D \big(\mc{E}(\ket{\Psi}_{\mathrm{mM}})\big) & = C_R\big(\mc{E}(\ket{\Psi}_{\mathrm{mM}})\big) \nonumber \\
		& = C_R(\ket{\Psi}_{\mathrm{mM}}) - C_\Delta(\ket{\Psi}_{\mathrm{mM}},A) \nonumber \\
		& = E_D(\ket{\Psi}_{\mathrm{mM}}) - C_\Delta(\ket{\Psi}_{\mathrm{mM}},A) \nonumber \\
		& \approx E_D(\ket{\Psi}_{\mathrm{mM}}) - \frac{h(\Delta^{-2})}{4} V(\ket{\Psi}_{\mathrm{mM}},A), 
\end{align}
where the final line uses the approximation (\ref{eqn:variance_approx2}) for a weak measurement.

(ii) A loss channel $\cL_\eta$ with ``efficiency'' $\eta$ (corresponding to the efficiency of the measurement device) models a process where each particle (subsystem) is lost to the environment with probability $1-\eta$, in other words the initial states of the particle and environment are eventually swapped. This symmetry implies that the transmitted state of the system $\cL_\eta(\rho)$ and the state of the partial state of environment $\rho^E_{\eta'}$ are the same if $\eta'=1-\eta$ (the roles of the ``transmitted'' and ``reflected'' systems are exchanged). For a family of states $\sS$ one defines $I_\eta(A:\ell)$ as the maximal MI  that is obtainable with a measurement device of efficiency $\eta$, and the corresponding $\t{MIC}_b(\sS)$ as the minimal efficiency that allows to obtain $b$ bits. Again, for the state $\ket{\Psi}_{\mathrm{mM}}$ the quantity $I_{1-\eta}(A:\ell)$ gives a bound on the decay of entanglement through the loss channel $\cL_{\eta}$, while $1-\t{MIC}_b$ is the minimal transmission of the channel that that leaves at least $E_F\big(\ket{\Psi}_{\mathrm{mM}}\big)-b$ bits of entanglement in the system.

\section{Conclusion and discussion}
\label{sec:conclusion}

Starting with the intuition that the macroscopic distinctness between two states ``dead and alive'' can be understood as ``the ease to distinguish'' the two states we lift this intuition to superpositions of multiple components by looking at ``the ease to obtain information'' about the state. We formalize this idea into a general measure that is also useful for mixed states. More precisely, we first quantify how much information one can extract from a pure state by measuring it with a classical detector with a limited resolution. Second, the minimal resolution that is required to extract the desired amount of information $b$ is associated with a measure that quantifies the ``macroscopic distinctness'' of the state, that we call macroscopicness of information content (MIC). Throughout a large part of the paper we use the Von Neumann model for a weak measurement of a fixed observable $A$ to model the classical detector.

To extend our measure to mixed states we use a convex roof construction, and illustrated it on a simple example.  It is argued that the parameter $b$ in our family of measures attributes a kind of ``macroscopicity rank'' to the superposition as it counts the effective number of components that are superposed. We also establish a relation between our measure and the variance of the state with respect to the opetator $A$ (its quantun Fisher information for a mixed state), that plays a central role in previously defined measures. In particular, we show that for a Gaussian pointer they are equal in the limit of small $b$.

Later, we present an alternative formulation of our measure, which stems from the same intuition but allows to directly deal with mixed stated without the detour of a heavy convex roof construction. It turns out to be equal to the basis-dependent quantum discord, and is also closely related to the measure for macroscopic distinctness that has been proposed in \cite{Kwon_Disturbance-Based_2016}. In particular, it also fulfills the proposed set of criteria for macroscopic coherence \cite{Yadin_General_2016}. So we can interpret is as the maximal scale at which the state provides the required amount of coherence, as quantified by $b$.  

Finally, we study the relation between the fragility of the state and its macroscopicness quantified by our measures. Concretely, we analyze the decay of the entanglement of formation in a micro-macro state when noise is applied on the macro side. We show that regardless of the model of the classical detector used to quantify the size, there is always a noise channel for which the fragility of entanglement is directly related to the macroscopicity of the state. This result is then applied to two models of classical detector: a weak measurement of $A$ central to the paper, and a generic inefficient detector modeled by a loss channel preceding an unknown measurement. 

Our work provides a novel tool to analyze and compare recent and future experiments aiming at the observation of quantum effects at larger and larger scales.

\textit{Acknowledgments.---} We thank Nicolas Sangouard for interesting discussions. This work was supported by the National Swiss Science Foundation (SNSF) projects  P2GEP2\_151964 and 00021\_149109,  and the Austrian Science Fund (FWF) P28000-N27, and the 
European Research Council (ERC MEC).
\appendix

\section{Details to equally-spaced-peaks example, Sec.~\ref{sec:exampl-equally-spac}}
\label{sec:deta-equally-spac}

The following calculation gives some details for the example discussed in Sec.~\ref{sec:exampl-equally-spac}. The probability of a measurement outcome $y$ is given 
\begin{equation}
p(y) = \frac{\#_\in^{(a, b]} }{2 r(k+1)},
\end{equation} 
where $\#_\in^{(a, b]}$ counts the number of peaks (elements of $\sS$) in the interval $(a, b]$. Its value can be expressed as the difference $\#_\in^{(a,b]} = \#_\leq^b- \#_\leq^a$ of the number of peaks with $y\leq b$ (respectively $y\leq a$), which reads
\begin{equation}
\#_\leq^y= 
\begin{cases}
0& y<0\\
\floor{y k}+1 & 0 \leq y <1 \\
k+1 & 1 \leq y 
\end{cases}
\end{equation}
with $\floor{x}$ denoting the integer part of $x$. The outcome $y$ can be equally well triggered by any peak from to the interval $(y-r, y+r]$, hence the conditional entropy is given by 
\begin{equation}
H(p(\ell|y))= \log_2\left(\#_\in^{(y-r, y+r]}\right).
\end{equation}
The MI reads
\begin{equation}
I(A:\ell) =\log_2(k+1) -  \int p(y) \log_2\left(\#_\in^{(y-r, y+r]}\right) dy 
\end{equation} 

Since both the probability of an outcome and the conditional entropy are uniquely determined by the number of peaks in the corresponding interval, one can rephrase the problem in terms of the random variable $n$ that englobes all the outcomes compatible with $n$ peaks. One has
\begin{equation}
P_n = \int\delta_{\#_\in^{(y-r, y+r]}}^n  p(y)  dy
\end{equation}
and
\begin{equation}
I(A:\ell) =\log_2(k+1) =\sum_{n=0}^{n_\t{max}} P_n \log_2(n).
\end{equation}

\section{Finite number of pure states in extremal ensemble}
\label{sec:finite-number-pure}

Here, we discuss a simplification in the convex-roof construction for measures defined for mixed states. Let us assume that $\rho$ is a full-rank state. If this is not true, one simply restricts the Hilbert space to the support of $\rho$. The number of ensemble averages of any non-pure density matrix $\rho$ is infinite. Moreover, even the number of pure states in such an ensemble is not bounded, one can even have ensembles defined by a non-discrete probability density on the manifold of pure states. This being said, the number of pure state in an \emph{extremal} ensemble is actually limited to $d^2$, where $d$ is the rank of $\rho$. This can be seen from the following argument.

First, there is a one-to-one correspondence between decompositions of $\rho$ (not necessarily in pure states) and partitions of identity (or POVMs) known as $\rho$-distortion \cite{Hughston_Complete_1993}. For each POVM, $\{E_i\}$ with $\sum_i E_i = \eins$, the operator $\rho_i= \frac{\sqrt{\rho}\, E_i \sqrt{\rho}}{\tr \rho E_i}$ is a valid state, and
\begin{equation}
 \sum_i \underbrace{p_i}_{\tr \rho E_i} \rho_i = \sqrt{\rho} \sum_i E_i \sqrt{\rho} =\rho.
\end{equation}
Second, $\rho_i$ is a pure state iff $E_i$ is rank one. This yields a one-to-one correspondence between ensemble partition (pure states) of $\rho$ and POVM composed of rank-one operators $\{E_i\}$. Finally, it is known that in dimension $d$ an extremal POVM, i.e. that a measurement that does not correspond to a mixture of different POVMs (such a procedure physically corresponds to randomly choosing the measurement to perform and forgetting the choice), has maximally $n=d^2$ elements \cite{Dariano_Classical_2005}. Via the correspondence above the same holds for extremal ensemble decomposition of $\rho$, and by construction the minimal size of Eq.~\eqref{eq:convex roof} is attained by an extremal ensemble.

\section{Convex roof example of Sec.~\ref{sec:mixed-states-convex}}
\label{app:roof}

\begin{figure}[!h]
\centering
\includegraphics[width=0.98\columnwidth]{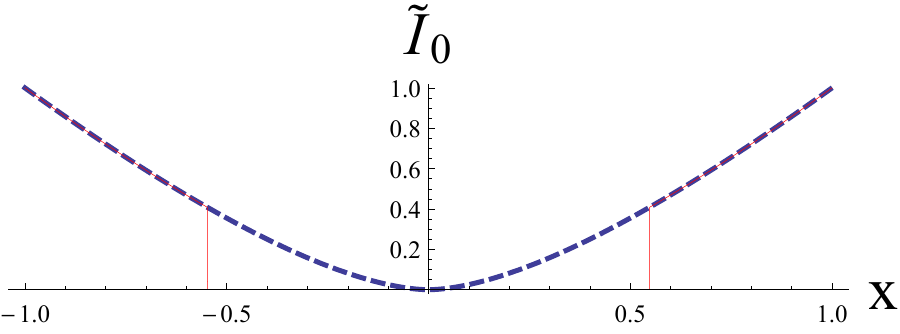}
\caption{Mutual information for pure states $I_0(x)$ (dashed, thick) and a rescaled size $\frac{b}{N} \t{MIC}_b$ for $b=0.4$ (solid thin) as functions of $x$.}
\label{fig:I0}
\end{figure}

\paragraph{Restriction to the XZ plane} Consider an ensemble decomposition $\sum_i q_i  \prjct{\Psi_i} = \sum_i \frac{q_i}{2}(\eins + {\bf r}_i\cdot {\bm \sigma})= \rho$. We show that there exists another decomposition that has a smaller or equal size but only involves states that lie in the XZ plane. To do so notice that
\begin{equation}
\sum q_i \bar \rho_i= \sum_i \frac{q_i}{2}(\eins + \bar {\bf r}_i \cdot {\bf \sigma}) =\rho
\end{equation}
 also holds for each $\bar {\bf r}_i  =(x_i\, 0\, z_i)$ restricted to the XZ plane. This is not a partition in pure state, but it naturally gives one, since each $\bar \rho_i$ can be decomposed in pure state as $\bar \rho_i = \lambda_i \prjct{\Psi_i^{(+)}  } +(1-\lambda_i) \prjct{\Psi_i^{(-)}}$ with the corresponding Bloch sphere vectors ${\bf r}_i^{(\pm)}= (x_i \,\, 0 \, \,\pm\sqrt{1-x_i^2})$. Finally, one has 
\begin{equation}
 \t{MIC}_b(\ket{\Psi_i}) \geq \lambda_i \t{MIC}_b(\ket{\Psi_i^{(+)}})+(1-\lambda_i) \t{MIC}_b(\ket{\Psi_i^{(-)}})
\end{equation}
since $|\sqrt{x_i^2+y_i^2}| \geq |x_i|$ and the size is monotonously increasing. Consequently the new decomposition
\begin{equation}
\rho = \sum_i q_i \sum_{s=\pm1} \big(1-(-1)^{s}(1-\lambda_i)\big) \prjct{\Psi_i^{(\t{sign}(s))}}
\end{equation}
yields a lower or equal size. It follows than from the beginning it is sufficient to only consider the ensembles where all elements lie in the XZ plane. As follows from [], extremal ensembles of this form involve three state at most.

\paragraph{Optimal ensemble}

Recall that the size of pure states in Eq.~\eqref{eq:size x} is zero for small $|x|$ (such that $\tilde I_0(x)< b$) and then increases monotonously with |$|x|$. In addition, $I_0(x)$ is convex in the regions $(r\equiv I_0^{-1}(b), 1]$ and $[-1, - r)$.

\begin{figure}[!ht]
\centering
\includegraphics[width=0.98\columnwidth]{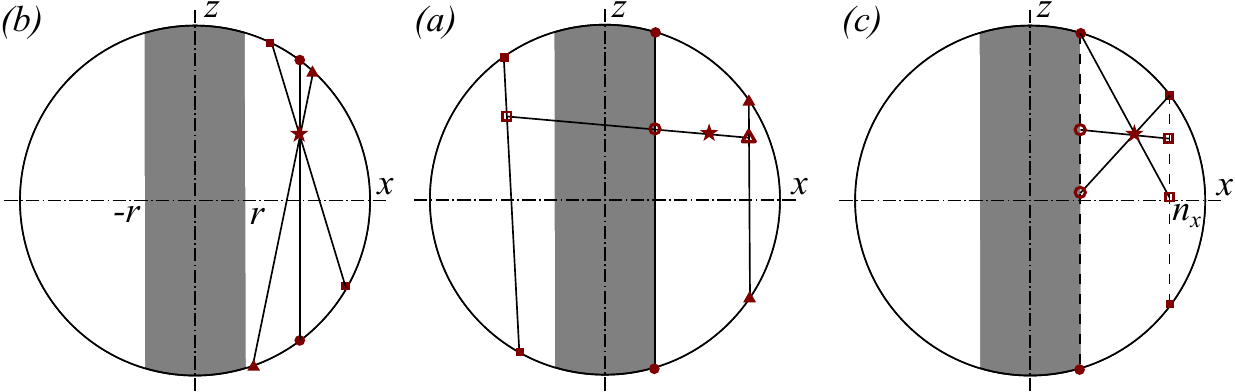}
\caption{A cut of the Bloch sphere through the XZ plane. The star represents the state $\rho$. (a) The convexity argument implies that vertical decomposition (circles) yields the lowest size among the three decomposition represented by line crossing $\rho$. (b) The ensemble consisting of states represented by triangles and circles always yields a lower size than the ensemble depicted by triangles and squares. (c) All the 3 ensembles give the same size.}
\label{fig:XZdec}
\end{figure}

If $|x_\rho|\leq r$, then the size of the state is zero. For example, this follows from the \emph{vertical} decomposition: $\rho$ is a mixture of the two pure states that have the same $x=x_\rho$, that both have zero size. So in the following we assume $x_\rho>r$. Without loss of generality, it follows that in the ensemble decomposition of $\rho$ there is at least one pure state that lies in the right white sector of the XZ circle on the right from $\rho$, see Fig.~\ref{fig:XZdec}. Actually, there are two possibilities: either (i) all the states lie in the right white sector, i.e. all these states satisfy $x \in (r,1]$, or (ii) some lie outside. 

The case (i) is rather simple, the convexity of the size in the $(r,1]$ region implies that the ensemble with the minimal average size should have all the pure states with the same $x$ as $\rho$, again this is fullfilled by the vertical decomposition, see Fig.~\ref{fig:XZdec}(a), and gives $\t{MIC}_b^{\t{vert.}}(\rho)=\t{MIC}_b(x_\rho)$.

The case (ii) is more involved, however it can be simplified by the following remark. Let us label all the pure in the ensemble states with $x\leq r$ by $\ket{\Phi_i}$ and all states the states with $x> r$ by $\ket{\Psi_i}$. We then have 
\begin{equation}
\rho = \underbrace{\sum_i q_i \prjct{\Psi_i}}_{\equiv p\, \sigma} + \underbrace{\sum_i \bar q_i \prjct{\Phi_i}}_{\equiv (1-p)\, \tau}, 
\end{equation}
with both $\sigma$ and $\tau$ valid density matrices, that are represented by the triangle and the empty square in Fig.~\ref{fig:XZdec}(b). The figure also directly suggests a decomposition that  has a smaller average size that the one we started with. This is given by
\begin{equation}
\rho = p' \sigma + (1-p')\tau',
\end{equation}
where $\tau'$ (represented by the emply circle in Fig.~\ref{fig:XZdec}(b))  lies on the intersection of the $x=r$ line and the line passing through $\tau$, $\rho$ and $\sigma$ and has a zero average size (think of its vertical decomposition).  In addition, one easily sees that $p'\leq p$ implying that this decomposition indeed yields a lower average size.

The two previous observation imply that minimal ensemble consists of at most of four pure state: the two states $\ket{\Psi_r^{(+)}}$ and $\ket{\Psi_r^{(-)}}$ with $x=r$ and the two states $\ket{\Psi_{n_x}^{(+)}}$ and $\ket{\Psi_{n_x}^{(-)}}$ with $x=n_x$, where $n_x\geq x_\rho$ \footnote{A careful reader might recall that four element ensembles on the plane are not extremal.}. In addition, all such ensembles have the same average size 
\begin{equation}
\langle\t{MIC}_b\rangle_{n_x}= (q_{n_x}^{(+)}+q_{n_x}^{(-)}) \t{MIC}_b(n_x)=\frac{x_\rho-r}{n_x-r} \t{MIC}_b(n_x)
\end{equation}
since the total weight
\begin{equation}
q_{n_x}\equiv q_{n_x}^{(+)}+q_{n_x}^{(-)} =\frac{x_\rho-r}{n_x-r}
\end{equation}
only depends on $n_x$, see Fig.~\ref{fig:XZdec}(c). Furthermore the case (i), discussed above, corresponds to the extremal case of $\langle\t{MIC}_b\rangle_{n_x=x_\rho}$ for which $q_{n_x}=1$. Finally, note that from above $n_x$ is bounded by $n_x \leq n_x^\t{max}$ with
\begin{widetext}
\begin{equation}
n_x^\t{max}(x_\rho, z_\rho,r) =
\left\{\begin{array}{c} 1 \qquad \t{for} \qquad z_\rho\leq (1-x_\rho)\sqrt{\frac{1+r}{1-r}}\qquad \t{otherwise}\\
\frac{2 r^2 x_{\rho } \left(x_{\rho }^2+z_{\rho }^2+1\right)+r \left(2 x_{\rho }^2 \left(\sqrt{1-r^2} z_{\rho }-3\right)+\left(z_{\rho }^2-1\right) \left(2 \sqrt{1-r^2} z_{\rho }+z_{\rho }^2+1\right)-x_{\rho }^4\right)-2 x_{\rho } \left(x_{\rho }^2 \left(\sqrt{1-r^2} z_{\rho }-1\right)+\left(z_{\rho }^2-1\right) \left(\sqrt{1-r^2} z_{\rho }+1\right)\right)}{2 z_{\rho }^2 \left(2 r^2-2 r x_{\rho }+x_{\rho }^2-1\right)+\left(-2 r x_{\rho }+x_{\rho }^2+1\right){}^2+z_{\rho }^4}
\end{array}\right.,
\end{equation}
\end{widetext}
as follows from a simple geometrical argument. This allows to write the size of $\rho$ from Eq.~\eqref{eq:convex roof} in the form
\begin{equation}
\widehat{\t{MIC}}_b (\rho) = \min_{n_x\in [x_\rho, n_x^\t{max}] }\frac{x_\rho-r}{n_x-r} \t{MIC}_b(n_x)
\end{equation} 
with $r =I_0^{-1}(b)$, which is a well-behaved function that can be easily computed.

\section{Weak Gaussian measurement for $I_\Delta(A:\ell)$} \label{app:I_weak_limit}
For the inequality (\ref{eqn:variance_ineq}), we calculate the relative entropy between an arbitrary distribution $p(x)$ and the Gaussian $g_\Delta(x - \bar{x})$, where $\bar{x} := \int \dd x\; x p(x)$. The relative entropy between two distributions $p(x),q(x)$ is defined as
\begin{equation}
	S(p(x) || q(x)) := \int \dd x\; p(x) \log p(x) - p(x) \log q(x).
\end{equation}
Thus,
\begin{align}
	S(p(x) || g_\Delta(x-\bar{x})) & = -H(p(x)) - \int \dd x\; p(x) \log g_\Delta(x - \bar{x}) \nonumber \\
	 & \hspace{-7em} = -H(p(x)) + \frac{1}{2} \log(2\pi \Delta^2) + \int \dd x \; p(x) \frac{(x-\bar{x})^2}{(2 \ln 2) \Delta^2} \nonumber \\
	 & \hspace{-7em} = -H(p(x)) + \frac{1}{2} \log(2\pi \Delta^2) + \frac{\expect{(x-\bar{x})^2}_{p(x)}}{(2 \ln 2) \Delta^2}.
\end{align}
Choosing $p(x) = \sum_\ell p_\ell g_\Delta(x-a_\ell)$ gives
\begin{align}
S(p(x)) || g_\Delta(x-\bar{x})) & = -H(p(x)) + \frac{1}{2} \log(2\pi \Delta^2) \nonumber \\
	& + \frac{V(\ket{\Psi},A) + \Delta^2}{(2 \ln 2)\Delta^2} \nonumber \\
	& = - I_\Delta(A:\ell) + \frac{V(\ket{\Psi},A)}{(2 \ln 2)\Delta^2},
\end{align}
where we have used the fact that $p(x)$ is a convolution of $p_\ell$ and $g_\Delta(x)$, under which the variance is additive. The inequality then follows from the non-negativity of the relative entropy.

To show (\ref{eqn:variance_approx}), we first prove the following useful result relating to classical statistics:
Let $A,\,X$ be random variables and $B := X + tA$. For sufficiently small $t$,
\begin{equation}
	H(B) = H(X) + \frac{t^2}{2 \ln 2} \mc{F}_c(X) V(A) + O(t^3),
\end{equation}
where $\mc{F}_c(X)$ is the classical Fisher information of $X$, and $V(A)$ is the variance of $A$.

Denote the density functions of $X,\,B$ by $g(x),\, p(x)$ respectively, and let $A$ have values $a_\ell$ with probabilities $p_\ell$. The classical Fisher information of $X$ is defined by
\begin{equation}
	\mc{F}_c(X) := \int \dd x\; \frac{g'(x)^2}{g(x)}.
\end{equation}
From the definition of $B$, we have
\begin{align}
	p(x) & = \sum_\ell p_\ell g(x- t a_\ell) \nonumber \\
	& \approx \sum_\ell p_\ell \left[ g(x) - t a_\ell g'(x) + \frac{t^2 a_\ell^2}{2} g''(x) \right] \nonumber \\
	& = g(x) - t \expect{A} g'(x) + \frac{t^2 \expect{A^2}}{2} g''(x),
\end{align}
where we have done an expansion to $O(t^2)$. Similarly, it is easily shown that
\begin{align}
	p(x) \ln p(x) & \approx g(x) \ln g(x) + t \left[ -\expect{A} g'(x) - \expect{A} g'(x) \ln g(x) \right] \nonumber \\
	&  + t^2 \left[ \frac{\expect{A}^2 g'(x)^2}{2g(x)} + \frac{\expect{A^2}}{2} g''(x) \nonumber \right. \\
	& \qquad + \left. \frac{\expect{A^2}}{2} g''(x) \ln g(x) \right].
\end{align}
In order to find $H(B)$, we integrate by parts
\begin{align}
	\int_{-\infty}^\infty \dd x\; g'(x) (1 + \ln g(x)) & = \Big[ g(x)(1 + \ln g(x)) \Big]_{-\infty}^\infty \nonumber \\
& - \int_{-\infty}^\infty \dd x\; g'(x) = 0, \\
	\int_{-\infty}^\infty \dd x\; g''(x) (1 + \ln g(x)) & = \Big[ g'(x) (1+\ln g(x)) \Big]_{-\infty}^\infty \nonumber \\
	& - \int_{-\infty}^\infty \dd x\; \frac{g'(x)^2}{g(x)} = -\mc{F}_c(X),
\end{align}
assuming $g$ is sufficiently regular that $\lim_{x \to \pm \infty} g'(x) \ln g(x) = 0$. Hence we have
\begin{align}
	- \int_{-\infty}^\infty \dd x\; p(x) \ln p(x) & \approx -\int_{-\infty}^\infty \dd x\; g(x) \ln g(x) \nonumber \\
	&  + \frac{t^2}{2} \mc{F}_c(X) ( \expect{A^2} - \expect{A}^2),
\end{align}
from which the result follows.

Now it can be verified that $I_\Delta(A:\ell)$ is unchanged under a simultaneous rescaling $\Delta \to \alpha \Delta,\, A \to \alpha A$. So $I_\Delta(A:\ell)$ in the limit of large $\Delta$ is the same as taking small $t$ in $H(tA + X) - \frac{1}{2} \log (2\pi e/ t^2)$, where $t = 1/\Delta$ and $X$ is a standard Gaussian of unit variance and zero mean. Applying the above result, we get
\begin{equation}
	I_\Delta(A:\ell) \approx \frac{t^2}{2 \ln 2} \mc{F}_c(X) V(\ket{\Psi},A),
\end{equation}
and $\mc{F}_c(X) = 1$. 

\section{Coherence criteria for $I_\Delta(A:\ell)$.} \label{app:I_criteria}
We need only check (C1) for pure states, as a mixed state has $I_\Delta(A:\ell)_\rho = 0$ if and only if there exists an ensemble decomposition with $I_\Delta(A:\ell)_{\ket{\Psi_k}} = 0 \, \forall k$. Now the concavity of the entropy tells us that $H(\sum_\ell p_\ell g_\Delta(x-a_\ell)) \geq \frac{1}{2} \log(2\pi e \Delta^2)$, with equality if and only if all the functions $g_\Delta(x-a_\ell)$ are the same, i.e.\ when $\ket{\Psi} = \ket{A_\ell}$ for some $\ell$.

(C3) follows immediately from the convex roof definition.

(C2b) can first be shown for pure states. Suppose that a stochastic free operation takes $\ket{\Psi} \to \ket{\Phi_\mu}$ with probability $w_\mu$, where $\sqrt{w_\mu} \ket{\Phi_\mu} = K_\mu \ket{\Psi}$, $K_\mu$ being a set of covariant Kraus operators, which take the form \cite{Yadin_General_2016}
\begin{equation}
	K_\mu = \sum_{i,j:\, a_i-a_j=\delta_\mu}  c^\mu_{i,j} \coh{A_i}{A_j}.
\end{equation}
Define $p_\ell$ as the probability of measuring $A=a_\ell$ for $\ket{\Psi}$, and $p^\mu_\ell$ as the probability of $a_\ell+\delta_\mu$ for $\ket{\Phi_\mu}$. Then it can be shown that $p_\ell = \sum_\mu w_\mu p^\mu_a$ -- this is because each $p^\mu_\ell$ distribution is obtained by measuring and shifting $p_\ell$ by $\delta_\mu$. Let $X$ be a standard Gaussian-distributed random variable. The distribution of $A + \Delta X$ for $\ket{\Psi}$ is $f(x) = \sum_\ell p_\ell g_\Delta(x-a_\ell)$, and similarly for $\ket{\Phi_\mu}$ we have $f_\mu(x) = \sum_\ell p^\mu_\ell g_\Delta(x-a_\ell)$. Now $f(x) = \sum_\mu w_\mu f_\mu(x)$ and so the concavity of the entropy gives $H(f) \geq \sum_\mu w_\mu H(f_\mu)$. It follows that $I_\Delta(A:\ell)_{\ket{\Psi}} \geq \sum_\mu w_\mu I_\Delta(A:\ell)_{\ket{\Phi_\mu}}$, as required.

(C2a,b) hold as a consequence of convexity and (C2b) for pure states -- see [ref] for the argument.

Finally, for (C4), we let $\ket{\Psi} = (\ket{A_i}+\ket{A_j})/\sqrt{2}$. It is clear that $I_\Delta(A:\ell)_{\ket{\Psi}}$ is a function of only $|a_i-a_j|$ and $\Delta$. So the requirement that $I_\Delta(A:\ell)_{\ket{\Psi}}$ be increasing with $|a_i-a_j|$ is equivalent to it being increasing under a replacement $A \to \alpha A$ with $\alpha > 1$. By the scale-invariance property mentioned in Appendix \ref{app:I_weak_limit}, this change of scale can be transferred to $\Delta \to \Delta / \alpha$. In fact, it should be clear that $I_\Delta(A:\ell)_{\ket{\Psi}}$ is decreasing with $\Delta$ (as more noise in the measurement cannot increase the mutual information) -- so this property holds. 

\section{Coherence criteria for $\widehat{\t{MIC}}'_b$}
For the vanishing criterion, note that either $I_0(A:\ell) \leq b$ or else there exists $\Delta > 0$ such that $I_\Delta(A:\ell) = b$ -- this follows from it being decreasing with $\Delta$ and from $I_\Delta(A:\ell) \to 0$ as $\Delta \to \infty$.

(C2a) for $I_\Delta(A:\ell)$ immediately implies the same for $\widehat{\t{MIC}}'_b$. This is because $I_\Delta(A:\ell)_{\mc{E}(\rho)} \leq I_\Delta(A:\ell)_\rho$ (for covariant $\mc{E}$) shows that the set of values of $\Delta$ for which $I_\Delta(A:\ell)_{\mc{E}(\rho)} \geq b$ is at most as big as the corresponding set for $\rho$.

Similarly, since $I_\Delta(A:\ell)$ is an increasing function of $|a_i-a_j|$ for $\ket{\Psi} = (\ket{A_i}+\ket{A_j})/\sqrt{2}$, the same is seen to be true for $\widehat{\t{MIC}}'_b$.

\section{Coherence criteria for $C_\Delta$} \label{app:C_criteria}
The channel taking induced by interacting $M$ with $P$, and then tracing out $P$, is denoted by $\Phi^\Delta$ -- we refer to this as partial dephasing. Note that the full dephasing operation is $\mathcal{G} = \Phi^0$. We first list some useful properties of the partial dephasing channel:
\begin{enumerate}
	\item[(i)] $\Phi^\Delta = \int \dd k\; f_\Delta(k) \mc{U}_k$, where $f_\Delta(k) = \sqrt{\frac{2}{\pi}}\Delta e^{-\Delta^2 k^2}$ is the momentum-space probability distribution for $\ket{\xi_\Delta}$, and $\mc{U}_k = e^{-ikA}$;
	\item[(ii)] $[\mc{E}, \Phi^\Delta] = 0$ for any covariant channel $\mc{E}$;
	\item[(iii)] $\Phi^\Delta(\coh{A_i}{A_j}) = e^{-\frac{(a_i-a_j)^2}{8 \Delta^2}} \coh{A_i}{A_j}$;
	\item[(iv)] $\Phi^\alpha \circ \Phi^\beta = \Phi^\gamma$, where $\gamma^{-2} = \alpha^{-2} + \beta^{-2}$.
\end{enumerate}
\begin{proof}
	For (i), we perform the partial trace over $P$ using momentum eigenstates $\ket{k}$:
	\begin{align}
\Phi^\Delta(\rho) & = \int \dd k\; {\bra{k}}_P e^{-iA \otimes \hat{p}} (\rho \otimes \prjct{\xi_\Delta}) e^{iA \otimes \hat{p}} {\ket{k}}_P \nonumber \\
	& = \int \dd k\; |\bracket{p}{\xi_\Delta}|^2 e^{-ik A} \rho e^{i k A} \nonumber \\
	& = \int \dd k\; f_\Delta(k) \mc{U}_k(\rho).
	\end{align}
	(ii) then immediately follows, since $[\mc{E},\mc{U}_k]=0$ by definition. Instead tracing out $P$ with position eigenstates, we have
	\begin{align}
\Phi^\Delta(\coh{A_i}{A_j}) & = \int \dd x\; {\bra{x}}_P e^{-iA \otimes \hat{p}} (\coh{A_i}{A_j} \otimes \prjct{\xi_\Delta} ) \nonumber \\
& \qquad \qquad e^{i A \otimes \hat{p}} {\ket{x}}_P \nonumber \\
		& = \coh{A_i}{A_j} \int \dd x\; \bra{x}{e^{-ia_i p}}\ket{\xi_\Delta} \bra{\xi_\Delta}{e^{ia_j p}}\ket{x} \nonumber \\
		& = \coh{A_i}{A_j} \int \dd x\; \bracket{x - a_i}{\xi_\Delta} \bracket{\xi_\Delta}{x - a_j} \nonumber \\
		& = \coh{A_i}{A_j} \int \dd x\; \frac{1}{\sqrt{2\pi}\Delta} e^{-\frac{1}{4\Delta^2}[(x-a_i)^2 + (x-a_j)^2]} \nonumber \\
		& = e^{-\frac{(a_i-a_j)^2}{8 \Delta^2}} \coh{A_i}{A_j},
	\end{align}
	showing (iii); part (iv) follows from this expression.
\end{proof}

In addition, $C_\Delta$ has the following properties:
\begin{enumerate}
	\item[(a)] $C_\Delta(\rho,A) \geq S(\rho || \Phi^{\Delta/\sqrt{2}}(\rho))$;
	\item[(b)] $C_\Delta(\rho,A) = \int \dd k\; f_\Delta(k) S(\mc{U}_k(\rho) || \Phi^\Delta(\rho))$;
	\item[(c)] decreasing with respect to $\Delta$;
	\item[(d)] invariant under a change of scale $A \to \alpha A,\, \Delta \to \alpha \Delta$.
\end{enumerate}
\begin{proof}
	(a) We need to use the fact that, for any quantum channel $\mc{N}$, $\tr[\mc{N}(\rho) \log \sigma] \leq \tr[\rho \log \mc{N}^\dagger(\sigma)]$, which is a consequence of the concavity of the logarithm \cite{Sutter_Strengthened_2016} (Lemma 3.6). From this, we have
	\begin{align}
		C_\Delta(\rho,A) & = -\tr[\Phi^\Delta(\rho) \log \Phi^\Delta(\rho)] - S(\rho) \nonumber\\
		& \geq -\tr[\rho \log \Phi^\Delta \circ \Phi^\Delta(\rho)] - S(\rho) \nonumber \\
		& = -\tr[\rho \log \Phi^{\Delta/\sqrt{2}}(\rho)] - S(\rho) \nonumber \\
		& = S(\rho || \Phi^{\Delta/\sqrt{2}}(\rho)),
	\end{align}
	having used property (iv) for the third line. \\
	
	(b) From property (i) above,
	\begin{align}
		C_\Delta(\rho,A) & = -\tr \left[ \int \dd k\; f_\Delta(k) \mc{U}_k(\rho) \log \Phi^\Delta(\rho) \right] - S(\rho) \nonumber \\
		& = \int \dd k\; f_\Delta(k) \left( -\tr[\mc{U}_k(\rho) \log \Phi^\Delta(\rho)] - S(\rho) \right) \nonumber \\
		& = \int \dd k\; f_\Delta(k) \left( - \tr[\mc{U}_k(\rho) \log \Phi^\Delta(\rho)] - S(\mc{U}_k(\rho)) \right) \nonumber \\
		& = \int \dd k\; f_\Delta(k) S(\mc{U}_k(\rho) || \Phi^\Delta(\rho)),
	\end{align}
	where for the third line, we have used the fact that $\mc{U}_k$ leaves the entropy unchanged. \\
	
	(c) Given some $\Delta_1 < \Delta_2$, there exists $\alpha \in (0,\infty)$ such that $\Delta_1^{-2} = \Delta_2^{-2} + \alpha^{-2}$. Then, by (iv) above, $\Phi^{\Delta_1} = \Phi^\alpha \circ \Phi^{\Delta_2}$. Therefore, taking $d$ as the Hilbert space dimension,
	\begin{align} \label{eqn:partial_dephasing_monotonicity}
		S(\Phi^{\Delta_1}(\rho) || I/d) & = S(\Phi^\alpha \circ \Phi^{\Delta_2}(\rho) || \Phi^\alpha(I/d)) \nonumber \\
			& \leq S(\Phi^{\Delta_2}(\rho)||I/d),
	\end{align}
	using the monotonicity of the relative entropy. Since $S(\rho || I/d) = \log d - S(\rho)$, this implies that $S(\Phi^{\Delta_1}(\rho)) \geq S(\Phi^{\Delta_2}(\rho))$. So $C_\Delta$ is decreasing with $\Delta$.

	(d) The scale-invariance is immediate from the fact that $\Phi^\Delta$ multiplies the matrix element $\coh{A_i}{A_j}$ by a function of $(a_i-a_j) / \Delta$.
\end{proof}

The properties of $\widetilde{\t{MIC}}_b$ follow exactly the same logic as for $\widehat{\t{MIC}}'_b$ above.
\section{Weak Gaussian measurement for $C_\Delta$} \label{app:C_weak_limit}
We do the calculation for a general state with rank $r$ strictly less than the dimension of the Hilbert space. We write the spectral decomposition $\rho = \sum_i \lambda_i \prjct{\psi_i}$, such that $\lambda_i > 0$ when $i < r$. Define the parameter $t = \Delta^{-2}$, which is assumed to be small. Let $\sigma_t = \Phi^{1/\sqrt{t}}(\rho) = \sum_i \mu_i \prjct{\phi_i}$, with its eigenvalues and eigenstates implicitly functions of $t$ and coinciding with those for $\rho$ at $t=0$. We also write their $t$-derivatives at $t=0$ as $\dot{\mu_i},\, \ket{\dot{\phi_i}}$. To lowest order, $\mu_i \approx \lambda_i + t \dot{\mu_i}$, so
\begin{align}
	S(\sigma_t) & = -\sum_i \mu_i \log \mu_i \nonumber \\
	& \approx -\sum _i (\lambda_i + t \dot{\mu_i}) \log (\lambda_i + t \dot{\mu_i}) \nonumber \\
	& = -\sum_{i<r} (\lambda_i + t\dot{\mu_i}) \left[ \log \lambda_i + \log(1 + t \dot{\mu_i}/\lambda_i) \right] \nonumber \\
	& \quad - \sum_{i \geq r} t \dot{\mu_i} \left[ \log t + \log \dot{\mu_i} \right].
\end{align}
After the constant term $- \sum_i \lambda_i \log \lambda_i$, the leading order is $O(t \log t)$, so $S(\sigma_t) \approx S(\rho) - t \log t \sum_{i \geq r} \dot{\mu_i}$. Now,
\begin{align}
	\bra{\psi_i}{\partial_t \sigma_t}\ket{\psi_i} & = \dot{\mu_i} + \lambda_i \big( \bracket{\psi_i}{\dot{\phi_i}} + \bracket{\dot{\phi_i}}{\psi_i} \big) \nonumber \\
	& = \dot{\mu_i} + \lambda_i \partial_t \bracket{\phi_i}{\phi_i} = \dot{\mu_i}.
\end{align}
Since $\sum_i \mu_i = 1$ is constant, we have $\sum_{i \geq r} \dot{\mu_i} = - \sum_{i < r} \dot{\mu_i} = -\sum_{i < r} \bra{\psi_i}{\partial_t \sigma_t}\ket{\psi_i} = - \tr(P_\rho \partial_t \sigma_t)$. 

It is easily verified that $\sigma_t$ evolves according to the master equation
\begin{equation}
	\partial_t \sigma_t = -\frac{1}{8} [A,[A,\sigma_t]].
\end{equation}
This can be seen by differentiating $\sigma_t = \Phi^{1/\sqrt{t}}(\rho)$ with respect to $t$, using property (iii) above of the dephasing channel. It follows that
\begin{align}
	\sum_{i \geq r} \dot{\mu_i} & = \frac{1}{8} \tr \left( P_\rho [A^2 \rho + \rho A^2 - 2 A \rho A] \right) \nonumber \\
	& = \frac{1}{4} \tr ( \rho A^2 - P_\rho A \rho A ),
\end{align}
having used the cyclic property of the trace and the fact that $P_\rho \rho = \rho P_\rho = \rho$. Putting these facts together gives the claimed result. For a pure state, $\rho = P_\rho = \prjct{\Psi}$, so then $\tr(P_\rho A \rho A) = \bra{\Psi}{A}\ket{\Psi}^2$.

\bibliographystyle{apsrev4-1}
\bibliography{References}

\end{document}